\documentclass[11pt,onecolumn]{IEEEtran}
\usepackage[usenames,dvipsnames]{xcolor}
\usepackage{amsmath}
\usepackage{amssymb}
\usepackage{amsfonts}
\usepackage{cite}
\usepackage{epsfig}
\usepackage{array}
\usepackage[font=small]{caption}
\usepackage{subfig}
\usepackage{pgf}
\usepackage{color}
\usepackage{algorithm}
\usepackage{float}
\usepackage{xspace}
\usepackage{enumerate}
\usepackage{wrapfig}
\usepackage[all]{xy}

\newtheorem{theorem}{Theorem}
\newtheorem{definition}{Definition}
\newtheorem{prop}[theorem]{Proposition}
\newtheorem{lemma}[theorem]{Lemma}

\newtheorem{remark}{Remark}
\newtheorem{example}{Example}

\newcommand{\beq}{\begin{equation}}
\newcommand{\eeq}{\end{equation}}
\newcommand{\bea}{\begin{eqnarray}}
\newcommand{\eea}{\end{eqnarray}}
\newcommand{\bean}{\begin{eqnarray*}}
\newcommand{\eean}{\end{eqnarray*}}
\newcommand{\bit}{\begin{itemize}}
\newcommand{\eit}{\end{itemize}}
\newcommand{\ben}{\begin{enumerate}}
\newcommand{\een}{\end{enumerate}}
\newcommand{\blem}{\begin{lem}}
\newcommand{\elem}{\end{lem}}
\newcommand{\bthm}{\begin{thm}}
\newcommand{\ethm}{\end{thm}}
\newcommand{\bpf}{\begin{IEEEproof}}
\newcommand{\epf}{\end{IEEEproof}}
\newcommand{\comment}[1]{}
\newcommand{\csota}{SMT-based solution\xspace}
\newcommand{\csotas}{SMT-based solution\xspace}
\newcommand{\sneak}{SNEAK\xspace}
\newcommand{\sneaka}{SNEAK\xspace}
\newcommand{\sotaabbr}{\textrm{SMT}}

\emergencystretch=\maxdimen 

\def\colornode{cyan!10!white}
\def\colordealer{cyan!5!white!92!purple}

\begin{document}

\title{Distributed Secret Dissemination Across a Network}
\author{Nihar~B.~Shah, K.~V.~Rashmi and Kannan~Ramchandran, {\em Fellow, IEEE}
\\Department of Electrical Engineering and Computer Sciences\\ University of California, Berkeley\\ \{nihar,\,rashmikv,\,kannanr\}@eecs.berkeley.edu
\thanks{This paper was presented, in part, at the IEEE International Symposium on Information Theory 2013~\cite{ourSecretSharingISIT}.}
\thanks{The work of Nihar B. Shah was supported by a Berkeley Fellowship and that of K. V. Rashmi was supported by a Facebook Fellowship.}
}

\maketitle
\vspace{-.4cm}
\thispagestyle{empty}

\begin{abstract}
Shamir's $(n, \, k)$ threshold secret sharing is an important component of several cryptographic protocols, such as those for secure multiparty-computation and key management. These protocols typically assume the presence of direct communication links from the dealer to all participants, in which case the dealer can directly pass the shares of the secret to each participant. In this paper, we consider the problem of secret sharing when the dealer does not have direct communication links to all the participants, and instead, the dealer and the participants form a general network. Existing methods are based on secure message transmissions from the dealer to each participant requiring considerable coordination in the network. 
In this paper, we present a distributed algorithm for disseminating shares over a network, which we call the \textit{\sneaka} algorithm, requiring each node to know only the identities of its one-hop neighbours. While \sneaka imposes a stronger condition on the network by requiring the dealer to be what we call $k$-propagating rather than $k$-connected as required by the existing solutions, we show that in addition to being distributed, \sneaka achieves significant reduction in the communication cost and the amount of randomness required. 


\end{abstract}


\section{Introduction}\label{sec:intro}
Shamir's classical $(n,\,k)$ secret sharing scheme~\cite{shamir1979share} is an essential ingredient of several cryptographic protocols. The scheme considers a set of $(n+1)$ entities: a \textit{dealer} and $n$ honest-but-curious \textit{participants}. The dealer possesses a secret $s$ and wishes to pass functions (called \textit{shares}) of this secret to the $n$ participants, such that the following two properties are satisfied:
\begin{itemize} 
\item \textit{$k$-secret-recovery}: the shares of any $k$ participants suffice to recover the secret,
\item \textit{$(k-1)$-collusion-resistance}: the aggregate data gathered by any $(k-1)$ nodes reveals no knowledge (in the information-theoretic sense) about the secret.
\end{itemize}

\newsavebox{\mynetwork}
\def\figToyHeight{1.1in}
\def\figToyHeightC{2in}
\def\needWidth{4.85in}
\def\needHeight{1.52in}
\def\nodeDia{11pt}
\def\nodegap{2.8}
\def\nodeAx{.2}
\def\nodeAy{6.7}

\def\nodeBx{\nodeAx+.35*\nodegap}
\def\nodeBy{\nodeAy-\nodegap}
\def\nodeCx{\nodeAx+.8*\nodegap}
\def\nodeCy{\nodeAy}
\def\nodeDx{\nodeBx+.9*\nodegap}
\def\nodeDy{\nodeBy}
\def\nodeEx{\nodeCx+.8*\nodegap}
\def\nodeEy{\nodeAy}
\def\nodeFx{\nodeDx+.8*\nodegap}
\def\nodeFy{\nodeBy}
\def\nodeDealerwidth{1.5*\nodeDia}
\def\nodeDealerheight{1.5*\nodeDia}
\def\nodeDealerx{\nodeAx-.6*\nodegap-.5*\nodeDealerwidth*2.5/72} 
\def\nodeDealery{\nodeBy+.65*\nodegap-.5*\nodeDealerheight*2.5/72)}
\def\nodeDealercornerx{\nodeDealerx-.5*\nodeDealerwidth*2.5/72}
\def\nodeDealercornery{\nodeDealery-.5*\nodeDealerheight*2.5/72}
\def\textedgeoffset{7pt}
\def\textedgeoffsetdown{-6pt}
\def\textedgeoffsetleft{9pt}
\def\textedgeoffsetleftdown{-6pt}
\def\textedgeoffsetright{6pt}
\def\textedgeoffsetrightdown{-9pt}

\def\vspacingBetnGraphs{.7cm}

\savebox{\mynetwork}{
\begin{pgfpicture}{0cm}{0cm}{\needWidth}{\needHeight}
\pgfline{\pgfxy(\nodeAx,\nodeAy)}{\pgfxy(\nodeCx,\nodeCy)}
\pgfline{\pgfxy(\nodeBx,\nodeBy)}{\pgfxy(\nodeDx,\nodeDy)}
\pgfline{\pgfxy(\nodeCx,\nodeCy)}{\pgfxy(\nodeEx,\nodeEy)}
\pgfline{\pgfxy(\nodeDx,\nodeDy)}{\pgfxy(\nodeFx,\nodeFy)}
\pgfline{\pgfxy(\nodeBx,\nodeBy)}{\pgfxy(\nodeCx,\nodeCy)}
\pgfline{\pgfxy(\nodeCx,\nodeCy)}{\pgfxy(\nodeDx,\nodeDy)}
\pgfline{\pgfxy(\nodeDx,\nodeDy)}{\pgfxy(\nodeEx,\nodeEy)}
\pgfline{\pgfxy(\nodeEx,\nodeEy)}{\pgfxy(\nodeFx,\nodeFy)}
\pgfline{\pgfxy(\nodeEx,\nodeEy)}{\pgfxy(\nodeFx,\nodeFy)}
\pgfline{\pgfxy(\nodeDealerx,\nodeDealery)}{\pgfxy(\nodeAx,\nodeAy)}
\pgfline{\pgfxy(\nodeDealerx,\nodeDealery)}{\pgfxy(\nodeBx,\nodeBy)}

\color{\colornode}
\pgfcircle[fill]{\pgfxy(\nodeAx,\nodeAy)}{\nodeDia}
\pgfcircle[fill]{\pgfxy(\nodeBx,\nodeBy)}{\nodeDia}
\pgfcircle[fill]{\pgfxy(\nodeCx,\nodeCy)}{\nodeDia}
\pgfcircle[fill]{\pgfxy(\nodeDx,\nodeDy)}{\nodeDia}
\pgfcircle[fill]{\pgfxy(\nodeEx,\nodeEy)}{\nodeDia}
\pgfcircle[fill]{\pgfxy(\nodeFx,\nodeFy)}{\nodeDia}
\color{\colordealer}
\pgfrect[fill]{\pgfxy(\nodeDealercornerx,\nodeDealercornery)}{\pgfpoint{\nodeDealerwidth}{\nodeDealerheight}}
\color{black}
\pgfcircle[stroke]{\pgfxy(\nodeAx,\nodeAy)}{\nodeDia}
\pgfcircle[stroke]{\pgfxy(\nodeBx,\nodeBy)}{\nodeDia}
\pgfcircle[stroke]{\pgfxy(\nodeCx,\nodeCy)}{\nodeDia}
\pgfcircle[stroke]{\pgfxy(\nodeDx,\nodeDy)}{\nodeDia}
\pgfcircle[stroke]{\pgfxy(\nodeEx,\nodeEy)}{\nodeDia}
\pgfcircle[stroke]{\pgfxy(\nodeFx,\nodeFy)}{\nodeDia}
\pgfrect[stroke]{\pgfxy(\nodeDealercornerx,\nodeDealercornery)}{\pgfpoint{\nodeDealerwidth}{\nodeDealerheight}}

\pgfputat{\pgfxy(\nodeAx,\nodeAy)}{\pgfbox[center,center]{\Large 1}}
\pgfputat{\pgfxy(\nodeBx,\nodeBy)}{\pgfbox[center,center]{\Large2}}
\pgfputat{\pgfxy(\nodeCx,\nodeCy)}{\pgfbox[center,center]{\Large3}}
\pgfputat{\pgfxy(\nodeDx,\nodeDy)}{\pgfbox[center,center]{\Large4}}
\pgfputat{\pgfxy(\nodeEx,\nodeEy)}{\pgfbox[center,center]{\Large5}}
\pgfputat{\pgfxy(\nodeFx,\nodeFy)}{\pgfbox[center,center]{\Large6}}
\pgfputat{\pgfxy(\nodeDealerx,\nodeDealery)}{\pgfbox[center,center]{\Large D}}
\end{pgfpicture}
}

\begin{figure*}[t]
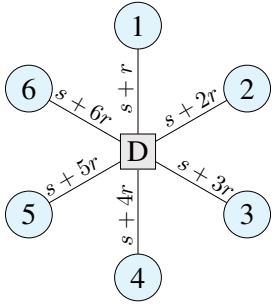
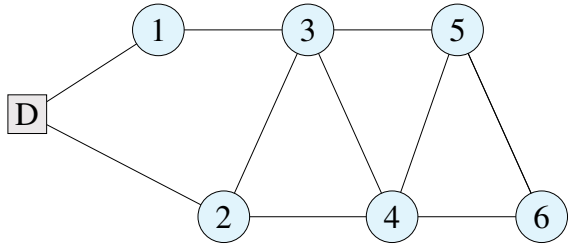

\centering
\subfloat[Dealer has communication links to all participants]{
\begin{minipage}[c]{.4\textwidth}
\vspace{2cm}
\hspace{-2cm}
\resizebox{1.44\textwidth}{!}{
\def\needWidth{2.6in}
\def\needHeight{2.6in}
\def\nodeDia{11pt}
\def\nodegap{2.8}

\def\radius{.82in}
\def\nodeDealerx{0.1in}
\def\nodeDealery{2.65in}
\def\nodeDealerwidth{1.5*\nodeDia}
\def\nodeDealerheight{1.5*\nodeDia}
\def\nodeDealercornerx{\nodeDealerx-.5*\nodeDealerwidth}
\def\nodeDealercornery{\nodeDealery-.5*\nodeDealerheight}

\def\nodeAx{\nodeDealerx}
\def\nodeAy{\nodeDealery + \radius}
\def\nodeBx{\nodeDealerx+.866*\radius}
\def\nodeBy{\nodeDealery+.5*\radius}
\def\nodeCx{\nodeDealerx+.866*\radius}
\def\nodeCy{\nodeDealery-.5*\radius}
\def\nodeDx{\nodeDealerx}
\def\nodeDy{\nodeDealery - \radius}
\def\nodeEx{\nodeDealerx-.866*\radius}
\def\nodeEy{\nodeDealery-.5*\radius}
\def\nodeFx{\nodeDealerx-.866*\radius}
\def\nodeFy{\nodeDealery+.5*\radius}

\begin{pgfpicture}{-3cm}{6cm}{\needWidth}{\needHeight}
\pgfline{\pgfpoint{\nodeDealerx}{\nodeDealery}}{\pgfpoint{\nodeAx}{\nodeAy}}
\pgfline{\pgfpoint{\nodeDealerx}{\nodeDealery}}{\pgfpoint{\nodeBx}{\nodeBy}}
\pgfline{\pgfpoint{\nodeDealerx}{\nodeDealery}}{\pgfpoint{\nodeCx}{\nodeCy}}
\pgfline{\pgfpoint{\nodeDealerx}{\nodeDealery}}{\pgfpoint{\nodeDx}{\nodeDy}}
\pgfline{\pgfpoint{\nodeDealerx}{\nodeDealery}}{\pgfpoint{\nodeEx}{\nodeEy}}
\pgfline{\pgfpoint{\nodeDealerx}{\nodeDealery}}{\pgfpoint{\nodeFx}{\nodeFy}}

\color{\colornode}
\pgfcircle[fill]{\pgfpoint{\nodeAx}{\nodeAy}}{\nodeDia}
\pgfcircle[fill]{\pgfpoint{\nodeBx}{\nodeBy}}{\nodeDia}
\pgfcircle[fill]{\pgfpoint{\nodeCx}{\nodeCy}}{\nodeDia}
\pgfcircle[fill]{\pgfpoint{\nodeDx}{\nodeDy}}{\nodeDia}
\pgfcircle[fill]{\pgfpoint{\nodeEx}{\nodeEy}}{\nodeDia}
\pgfcircle[fill]{\pgfpoint{\nodeFx}{\nodeFy}}{\nodeDia}

\color{\colordealer}
\pgfrect[fill]{\pgfpoint{\nodeDealercornerx}{\nodeDealercornery}}{\pgfpoint{\nodeDealerwidth}{\nodeDealerheight}}

\color{black}
\pgfcircle[stroke]{\pgfpoint{\nodeAx}{\nodeAy}}{\nodeDia}
\pgfcircle[stroke]{\pgfpoint{\nodeBx}{\nodeBy}}{\nodeDia}
\pgfcircle[stroke]{\pgfpoint{\nodeCx}{\nodeCy}}{\nodeDia}
\pgfcircle[stroke]{\pgfpoint{\nodeDx}{\nodeDy}}{\nodeDia}
\pgfcircle[stroke]{\pgfpoint{\nodeEx}{\nodeEy}}{\nodeDia}
\pgfcircle[stroke]{\pgfpoint{\nodeFx}{\nodeFy}}{\nodeDia}
\pgfrect[stroke]{\pgfpoint{\nodeDealercornerx}{\nodeDealercornery}}{\pgfpoint{\nodeDealerwidth}{\nodeDealerheight}}

\pgfputat{\pgfpoint{\nodeAx}{\nodeAy}}{\pgfbox[center,center]{\Large 1}}
\pgfputat{\pgfpoint{\nodeBx}{\nodeBy}}{\pgfbox[center,center]{\Large 2}}
\pgfputat{\pgfpoint{\nodeCx}{\nodeCy}}{\pgfbox[center,center]{\Large 3}}
\pgfputat{\pgfpoint{\nodeDx}{\nodeDy}}{\pgfbox[center,center]{\Large 4}}
\pgfputat{\pgfpoint{\nodeEx}{\nodeEy}}{\pgfbox[center,center]{\Large 5}}
\pgfputat{\pgfpoint{\nodeFx}{\nodeFy}}{\pgfbox[center,center]{\Large 6}}
\pgfputat{\pgfpoint{\nodeDealerx}{\nodeDealery}}{\pgfbox[center,center]{\Large D}}

\pgfputlabelrotated{.5}{\pgfpoint{\nodeDealerx}{\nodeDealery}}{\pgfpoint{\nodeAx}{\nodeAy}}{.2cm}{\pgfbox[center,center]{$s+r$}}
\pgfputlabelrotated{.54}{\pgfpoint{\nodeDealerx}{\nodeDealery}}{\pgfpoint{\nodeBx}{\nodeBy}}{.2cm}{\pgfbox[center,center]{$s+2r$}}
\pgfputlabelrotated{.57}{\pgfpoint{\nodeDealerx}{\nodeDealery}}{\pgfpoint{\nodeCx}{\nodeCy}}{.2cm}{\pgfbox[center,center]{$s+3r$}}
\pgfputlabelrotated{.5}{\pgfpoint{\nodeDx}{\nodeDy}}{\pgfpoint{\nodeDealerx}{\nodeDealery}}{.2cm}{\pgfbox[center,center]{$s+4r$}}
\pgfputlabelrotated{.45}{\pgfpoint{\nodeEx}{\nodeEy}}{\pgfpoint{\nodeDealerx}{\nodeDealery}}{.2cm}{\pgfbox[center,center]{$s+5r$}}
\pgfputlabelrotated{.45}{\pgfpoint{\nodeFx}{\nodeFy}}{\pgfpoint{\nodeDealerx}{\nodeDealery}}{.2cm}{\pgfbox[center,center]{$s+6r$}}
\end{pgfpicture}
}
\vspace{1.3cm}
\end{minipage}
\label{fig:fully_connected}
}
\subfloat[Dealer and participants form a general network]{
\begin{minipage}[c]{.53\textwidth}
\vspace{2.3cm}
\resizebox{1.44\textwidth}{!}{
\centering
\vspace{3cm}
\begin{pgfpicture}{-3cm}{3cm}{\needWidth}{\needHeight}
\usebox{\mynetwork}
\end{pgfpicture}
\vspace{1cm}
}
\end{minipage}
\label{fig:toy_a}
}
\caption{Shamir's secret sharing for $k=2$ and $n=6$. (a) All participants $(1,\ldots,6)$ are connected directly to the dealer $(D)$, allowing the dealer to directly pass the shares. The share of participant $i~(1 \leq i \leq 6)$ is $s+ir$, where $s$ is the secret and $r$ is a value chosen uniformly at random from the finite field of operation $\mathbb{F}_7$. (b) The dealer and the participants form a general network, where the dealer cannot pass shares directly to participants $3,\,4,\,5$ and $6$.
}
\label{fig:networks}
\end{figure*}

\begin{figure*}
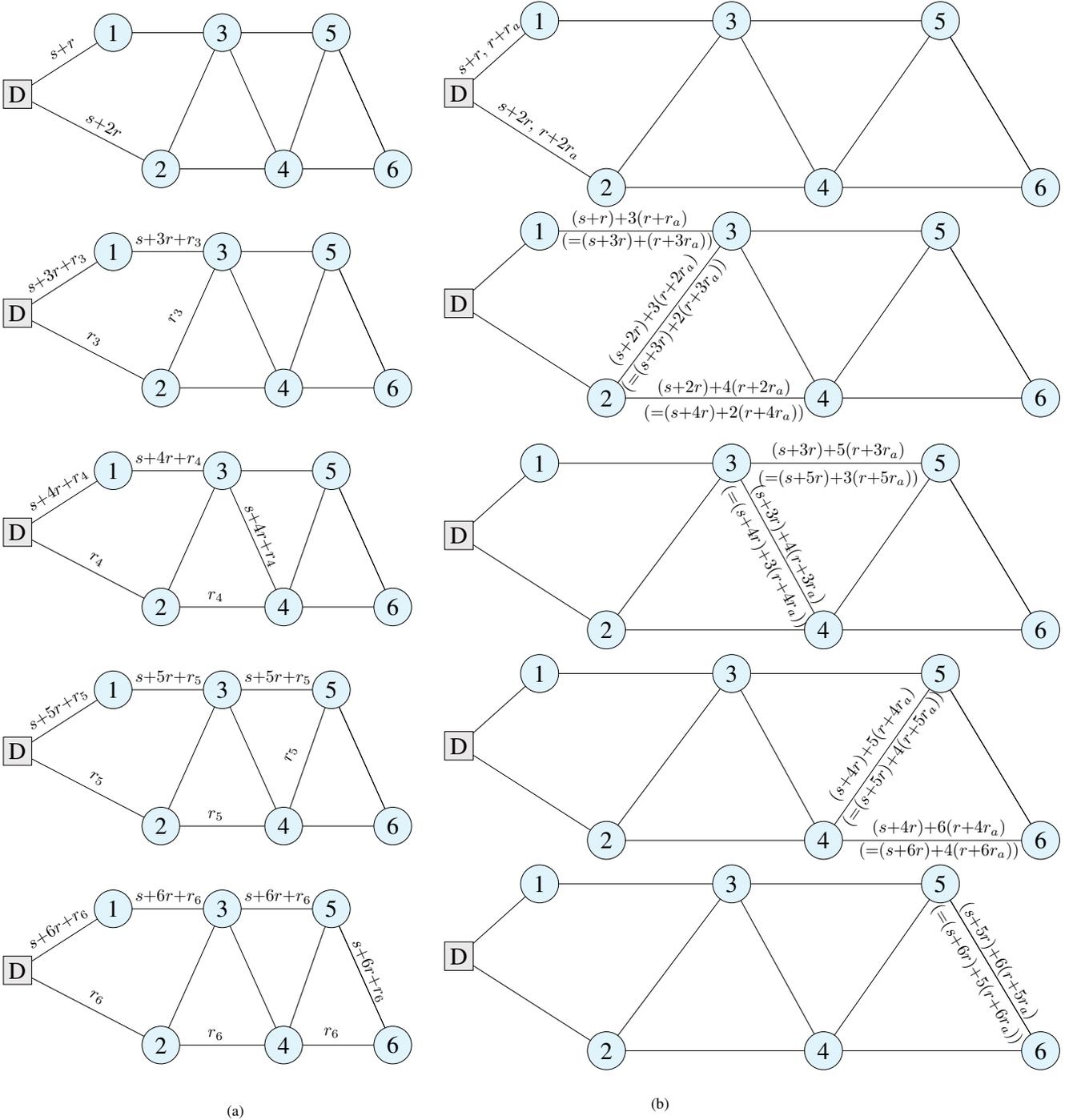

\vspace{-.5cm}
\thinmuskip=0\thinmuskip
\medmuskip=0\medmuskip
\thickmuskip=0\thickmuskip
\small

\def\needWidth{4.85in}
\def\needHeight{1.52in}
\def\figToyHeight{1.1in}
\def\figToyHeightC{2in}
\def\nodeDia{11pt}
\def\nodegap{2.8}
\def\nodeAx{.2}
\def\nodeAy{6.7}

\def\nodeBx{\nodeAx+.35*\nodegap}
\def\nodeBy{\nodeAy-\nodegap}
\def\nodeCx{\nodeAx+.8*\nodegap}
\def\nodeCy{\nodeAy}
\def\nodeDx{\nodeBx+.9*\nodegap}
\def\nodeDy{\nodeBy}
\def\nodeEx{\nodeCx+.8*\nodegap}
\def\nodeEy{\nodeAy}
\def\nodeFx{\nodeDx+.8*\nodegap}
\def\nodeFy{\nodeBy}
\def\nodeDealerwidth{1.5*\nodeDia}
\def\nodeDealerheight{1.5*\nodeDia}
\def\nodeDealerx{\nodeAx-.6*\nodegap-.5*\nodeDealerwidth*2.5/72} 
\def\nodeDealery{\nodeBy+.65*\nodegap-.5*\nodeDealerheight*2.5/72)}
\def\nodeDealercornerx{\nodeDealerx-.5*\nodeDealerwidth*2.5/72}
\def\nodeDealercornery{\nodeDealery-.5*\nodeDealerheight*2.5/72}
\def\textedgeoffset{7pt}
\def\textedgeoffsetdown{-6pt}
\def\textedgeoffsetleft{9pt}
\def\textedgeoffsetleftdown{-6pt}
\def\textedgeoffsetright{6pt}
\def\textedgeoffsetrightdown{-9pt}
\def\colornode{cyan!10!white}
\def\colordealer{cyan!5!white!92!purple}

\def\vspacingBetnGraphs{.2cm}

\hspace{1.3cm}
\begin{minipage}[b]{.32\textwidth}
\resizebox{.86\textwidth}{!}{
\subfloat[~]{
\begin{minipage}[l]{\textwidth}

\begin{pgfpicture}{0cm}{0cm}{\needWidth}{\needHeight}
\pgfputlabelrotated{.6}{\pgfxy(\nodeDealerx,\nodeDealery)}{\pgfxy(\nodeAx,\nodeAy)}{\textedgeoffsetright}{\pgfbox[center,center]{$s+r$}}
\pgfputlabelrotated{.6}{\pgfxy(\nodeDealerx,\nodeDealery)}{\pgfxy(\nodeBx,\nodeBy)}{\textedgeoffsetleft}{\pgfbox[center,center]{$s+2r$}}
\usebox{\mynetwork}
\end{pgfpicture}
\vspace{\vspacingBetnGraphs}

\begin{pgfpicture}{0cm}{0cm}{\needWidth}{\needHeight}
\def\patha{$s+3r+r_3$}
\def\pathb{$r_3$}
\pgfputlabelrotated{.53}{\pgfxy(\nodeDealerx,\nodeDealery)}{\pgfxy(\nodeAx,\nodeAy)}{\textedgeoffsetright}{\pgfbox[center,center]{\patha}}
\pgfputlabelrotated{.53}{\pgfxy(\nodeDealerx,\nodeDealery)}{\pgfxy(\nodeBx,\nodeBy)}{\textedgeoffsetleft}{\pgfbox[center,center]{$r_3$}}
\pgfputlabelrotated{.57}{\pgfxy(\nodeAx,\nodeAy)}{\pgfxy(\nodeCx,\nodeCy)}{\textedgeoffset}{\pgfbox[center,center]{$s+3r+r_3$}}
\pgfputlabelrotated{.5}{\pgfxy(\nodeBx,\nodeBy)}{\pgfxy(\nodeCx,\nodeCy)}{\textedgeoffsetright}{\pgfbox[center,center]{$r_3$}}
\usebox{\mynetwork}
\end{pgfpicture}
\vspace{\vspacingBetnGraphs}

\begin{pgfpicture}{0cm}{0cm}{\needWidth}{\needHeight}
\def\patha{$s+4r+r_4$}
\def\pathb{$r_4$}
\pgfputlabelrotated{.55}{\pgfxy(\nodeDealerx,\nodeDealery)}{\pgfxy(\nodeAx,\nodeAy)}{\textedgeoffsetright}{\pgfbox[center,center]{\patha}}
\pgfputlabelrotated{.55}{\pgfxy(\nodeDealerx,\nodeDealery)}{\pgfxy(\nodeBx,\nodeBy)}{\textedgeoffsetleft}{\pgfbox[center,center]{\pathb}}
\pgfputlabelrotated{.57}{\pgfxy(\nodeAx,\nodeAy)}{\pgfxy(\nodeCx,\nodeCy)}{\textedgeoffset}{\pgfbox[center,center]{\patha}}
\pgfputlabelrotated{.5}{\pgfxy(\nodeBx,\nodeBy)}{\pgfxy(\nodeDx,\nodeDy)}{\textedgeoffsetright}{\pgfbox[center,center]{\pathb}}
\pgfputlabelrotated{.52}{\pgfxy(\nodeCx,\nodeCy)}{\pgfxy(\nodeDx,\nodeDy)}{\textedgeoffsetleft}{\pgfbox[center,center]{\patha}}
\usebox{\mynetwork}
\end{pgfpicture}
\vspace{\vspacingBetnGraphs}

\begin{pgfpicture}{0cm}{0cm}{\needWidth}{\needHeight}
\def\patha{$s+5r+r_5$}
\def\pathb{$r_5$}
\pgfputlabelrotated{.55}{\pgfxy(\nodeDealerx,\nodeDealery)}{\pgfxy(\nodeAx,\nodeAy)}{\textedgeoffsetright}{\pgfbox[center,center]{\patha}}
\pgfputlabelrotated{.55}{\pgfxy(\nodeDealerx,\nodeDealery)}{\pgfxy(\nodeBx,\nodeBy)}{\textedgeoffsetleft}{\pgfbox[center,center]{\pathb}}
\pgfputlabelrotated{.57}{\pgfxy(\nodeAx,\nodeAy)}{\pgfxy(\nodeCx,\nodeCy)}{\textedgeoffset}{\pgfbox[center,center]{\patha}}
\pgfputlabelrotated{.5}{\pgfxy(\nodeBx,\nodeBy)}{\pgfxy(\nodeDx,\nodeDy)}{\textedgeoffsetright}{\pgfbox[center,center]{\pathb}}
\pgfputlabelrotated{.57}{\pgfxy(\nodeCx,\nodeCy)}{\pgfxy(\nodeEx,\nodeEy)}{\textedgeoffset}{\pgfbox[center,center]{\patha}}
\pgfputlabelrotated{.5}{\pgfxy(\nodeDx,\nodeDy)}{\pgfxy(\nodeEx,\nodeEy)}{\textedgeoffsetright}{\pgfbox[center,center]{\pathb}}
\usebox{\mynetwork}
\end{pgfpicture}
\vspace{\vspacingBetnGraphs}

\begin{pgfpicture}{0cm}{0cm}{\needWidth}{\needHeight}
\def\patha{$s+6r+r_6$}
\def\pathb{$r_6$}
\pgfputlabelrotated{.55}{\pgfxy(\nodeDealerx,\nodeDealery)}{\pgfxy(\nodeAx,\nodeAy)}{\textedgeoffsetright}{\pgfbox[center,center]{\patha}}
\pgfputlabelrotated{.55}{\pgfxy(\nodeDealerx,\nodeDealery)}{\pgfxy(\nodeBx,\nodeBy)}{\textedgeoffsetleft}{\pgfbox[center,center]{\pathb}}
\pgfputlabelrotated{.57}{\pgfxy(\nodeAx,\nodeAy)}{\pgfxy(\nodeCx,\nodeCy)}{\textedgeoffset}{\pgfbox[center,center]{\patha}}
\pgfputlabelrotated{.5}{\pgfxy(\nodeBx,\nodeBy)}{\pgfxy(\nodeDx,\nodeDy)}{\textedgeoffsetright}{\pgfbox[center,center]{\pathb}}
\pgfputlabelrotated{.57}{\pgfxy(\nodeCx,\nodeCy)}{\pgfxy(\nodeEx,\nodeEy)}{\textedgeoffset}{\pgfbox[center,center]{\patha}}
\pgfputlabelrotated{.5}{\pgfxy(\nodeDx,\nodeDy)}{\pgfxy(\nodeFx,\nodeFy)}{\textedgeoffset}{\pgfbox[center,center]{\pathb}}
\pgfputlabelrotated{.5}{\pgfxy(\nodeEx,\nodeEy)}{\pgfxy(\nodeFx,\nodeFy)}{\textedgeoffsetleft}{\pgfbox[center,center]{\patha}}
\usebox{\mynetwork}
\end{pgfpicture}
\vspace{-1.3in}
\label{fig:toy_b}
\end{minipage}
}}
\end{minipage}
\;\hspace{-.3cm}
\begin{minipage}[b]{.48\textwidth}
\vspace{2.3cm}
\def\needWidth{6in}
\def\needHeight{2.2in}
\resizebox{.85\textwidth}{!}{
\subfloat[~]{
\begin{minipage}[l]{\textwidth}
\def\needWidth{6in}
\def\needHeight{2.2in}
\def\nodeDia{11pt}
\def\nodegap{3.4}
\def\nodeAx{1.72}
\def\nodeAy{8.8}

\def\nodeBx{\nodeAx+.4*\nodegap}
\def\nodeBy{\nodeAy-\nodegap}
\def\nodeCx{\nodeAx+1.15*\nodegap}
\def\nodeCy{\nodeAy}
\def\nodeDx{\nodeBx+1.3*\nodegap}
\def\nodeDy{\nodeBy}
\def\nodeEx{\nodeCx+1.25*\nodegap}
\def\nodeEy{\nodeAy}
\def\nodeFx{\nodeDx+1.3*\nodegap}
\def\nodeFy{\nodeBy}
\def\nodeDealerwidth{1.5*\nodeDia}
\def\nodeDealerheight{1.5*\nodeDia}
\def\nodeDealerx{\nodeAx-.4*\nodegap-.5*\nodeDealerwidth*2.5/72} 
\def\nodeDealery{\nodeBy+.65*\nodegap-.5*\nodeDealerheight*2.5/72)}
\def\nodeDealercornerx{\nodeDealerx-.5*\nodeDealerwidth*2.5/72}
\def\nodeDealercornery{\nodeDealery-.5*\nodeDealerheight*2.5/72}
\def\textedgeoffset{7pt}
\def\textedgeoffsetdown{-6pt}
\def\textedgeoffsetleft{9pt}
\def\textedgeoffsetleftdown{-6pt}
\def\textedgeoffsetright{6pt}
\def\textedgeoffsetrightdown{-9pt}

\def\vspacingBetnGraphs{-1.3cm}

\newsavebox{\mynetworkC}
\savebox{\mynetworkC}{
\begin{pgfpicture}{0cm}{0cm}{\needWidth}{\needHeight}
\pgfline{\pgfxy(\nodeAx,\nodeAy)}{\pgfxy(\nodeCx,\nodeCy)}
\pgfline{\pgfxy(\nodeBx,\nodeBy)}{\pgfxy(\nodeDx,\nodeDy)}
\pgfline{\pgfxy(\nodeCx,\nodeCy)}{\pgfxy(\nodeEx,\nodeEy)}
\pgfline{\pgfxy(\nodeDx,\nodeDy)}{\pgfxy(\nodeFx,\nodeFy)}
\pgfline{\pgfxy(\nodeBx,\nodeBy)}{\pgfxy(\nodeCx,\nodeCy)}
\pgfline{\pgfxy(\nodeCx,\nodeCy)}{\pgfxy(\nodeDx,\nodeDy)}
\pgfline{\pgfxy(\nodeDx,\nodeDy)}{\pgfxy(\nodeEx,\nodeEy)}
\pgfline{\pgfxy(\nodeEx,\nodeEy)}{\pgfxy(\nodeFx,\nodeFy)}
\pgfline{\pgfxy(\nodeEx,\nodeEy)}{\pgfxy(\nodeFx,\nodeFy)}
\pgfline{\pgfxy(\nodeDealerx,\nodeDealery)}{\pgfxy(\nodeAx,\nodeAy)}
\pgfline{\pgfxy(\nodeDealerx,\nodeDealery)}{\pgfxy(\nodeBx,\nodeBy)}

\color{\colornode}
\pgfcircle[fill]{\pgfxy(\nodeAx,\nodeAy)}{\nodeDia}
\pgfcircle[fill]{\pgfxy(\nodeBx,\nodeBy)}{\nodeDia}
\pgfcircle[fill]{\pgfxy(\nodeCx,\nodeCy)}{\nodeDia}
\pgfcircle[fill]{\pgfxy(\nodeDx,\nodeDy)}{\nodeDia}
\pgfcircle[fill]{\pgfxy(\nodeEx,\nodeEy)}{\nodeDia}
\pgfcircle[fill]{\pgfxy(\nodeFx,\nodeFy)}{\nodeDia}
\color{\colordealer}
\pgfrect[fill]{\pgfxy(\nodeDealercornerx,\nodeDealercornery)}{\pgfpoint{\nodeDealerwidth}{\nodeDealerheight}}
\color{black}
\pgfcircle[stroke]{\pgfxy(\nodeAx,\nodeAy)}{\nodeDia}
\pgfcircle[stroke]{\pgfxy(\nodeBx,\nodeBy)}{\nodeDia}
\pgfcircle[stroke]{\pgfxy(\nodeCx,\nodeCy)}{\nodeDia}
\pgfcircle[stroke]{\pgfxy(\nodeDx,\nodeDy)}{\nodeDia}
\pgfcircle[stroke]{\pgfxy(\nodeEx,\nodeEy)}{\nodeDia}
\pgfcircle[stroke]{\pgfxy(\nodeFx,\nodeFy)}{\nodeDia}
\pgfrect[stroke]{\pgfxy(\nodeDealercornerx,\nodeDealercornery)}{\pgfpoint{\nodeDealerwidth}{\nodeDealerheight}}

\pgfputat{\pgfxy(\nodeAx,\nodeAy)}{\pgfbox[center,center]{\Large 1}}
\pgfputat{\pgfxy(\nodeBx,\nodeBy)}{\pgfbox[center,center]{\Large2}}
\pgfputat{\pgfxy(\nodeCx,\nodeCy)}{\pgfbox[center,center]{\Large3}}
\pgfputat{\pgfxy(\nodeDx,\nodeDy)}{\pgfbox[center,center]{\Large4}}
\pgfputat{\pgfxy(\nodeEx,\nodeEy)}{\pgfbox[center,center]{\Large5}}
\pgfputat{\pgfxy(\nodeFx,\nodeFy)}{\pgfbox[center,center]{\Large6}}
\pgfputat{\pgfxy(\nodeDealerx,\nodeDealery)}{\pgfbox[center,center]{\Large D}}
\end{pgfpicture}
}


\begin{pgfpicture}{0cm}{0cm}{\needWidth}{\needHeight}
\pgfputlabelrotated{.53}{\pgfxy(\nodeDealerx,\nodeDealery)}{\pgfxy(\nodeAx,\nodeAy)}{\textedgeoffsetright}{\pgfbox[center,center]{$s+r,~r+r_a$}}
\pgfputlabelrotated{.53}{\pgfxy(\nodeDealerx,\nodeDealery)}{\pgfxy(\nodeBx,\nodeBy)}{\textedgeoffsetleft}{\pgfbox[center,center]{$s+2r,~r+2r_a$}}
\usebox{\mynetworkC}
\end{pgfpicture}\vspace{\vspacingBetnGraphs}

\begin{pgfpicture}{0cm}{0cm}{\needWidth}{\needHeight}
\pgfputlabelrotated{.5}{\pgfxy(\nodeAx,\nodeAy)}{\pgfxy(\nodeCx,\nodeCy)}{\textedgeoffset}{\pgfbox[center,center]{$(s+r)+3(r+r_a)$}}
\pgfputlabelrotated{.54}{\pgfxy(\nodeAx,\nodeAy)}{\pgfxy(\nodeCx,\nodeCy)}{\textedgeoffsetdown}{\pgfbox[center,center]{$\left(=(s+3r)+(r+3r_a)\right)$}}
\pgfputlabelrotated{.49}{\pgfxy(\nodeBx,\nodeBy)}{\pgfxy(\nodeCx,\nodeCy)}{\textedgeoffsetright}{\pgfbox[center,center]{$(s+2r)+3(r+2r_a)$}}
\pgfputlabelrotated{.49}{\pgfxy(\nodeBx,\nodeBy)}{\pgfxy(\nodeCx,\nodeCy)}{\textedgeoffsetrightdown}{\pgfbox[center,center]{$\left(=(s+3r)+2(r+3r_a)\right)$}}
\pgfputlabelrotated{.57}{\pgfxy(\nodeBx,\nodeBy)}{\pgfxy(\nodeDx,\nodeDy)}{\textedgeoffsetright}{\pgfbox[center,center]{$(s+2r)+4(r+2r_a)$}}
\pgfputlabelrotated{.57}{\pgfxy(\nodeBx,\nodeBy)}{\pgfxy(\nodeDx,\nodeDy)}{\textedgeoffsetrightdown}{\pgfbox[center,center]{$\left(=(s+4r)+2(r+4r_a)\right)$}}
\usebox{\mynetworkC}
\end{pgfpicture}
\vspace{\vspacingBetnGraphs}

\begin{pgfpicture}{0cm}{0cm}{\needWidth}{\needHeight}
\pgfputlabelrotated{.53}{\pgfxy(\nodeCx,\nodeCy)}{\pgfxy(\nodeDx,\nodeDy)}{\textedgeoffsetleft}{\pgfbox[center,center]{$(s+3r)+4(r+3r_a)$}}
\pgfputlabelrotated{.53}{\pgfxy(\nodeCx,\nodeCy)}{\pgfxy(\nodeDx,\nodeDy)}{\textedgeoffsetleftdown}{\pgfbox[center,center]{$\left(=(s+4r)+3(r+4r_a)\right)$}}
\pgfputlabelrotated{.54}{\pgfxy(\nodeCx,\nodeCy)}{\pgfxy(\nodeEx,\nodeEy)}{\textedgeoffset}{\pgfbox[center,center]{$(s+3r)+5(r+3r_a)$}}
\pgfputlabelrotated{.54}{\pgfxy(\nodeCx,\nodeCy)}{\pgfxy(\nodeEx,\nodeEy)}{\textedgeoffsetrightdown}{\pgfbox[center,center]{$\left(=(s+5r)+3(r+5r_a)\right)$}}
\usebox{\mynetworkC}
\end{pgfpicture}\vspace{\vspacingBetnGraphs}

\begin{pgfpicture}{0cm}{0cm}{\needWidth}{\needHeight}
\pgfputlabelrotated{.54}{\pgfxy(\nodeDx,\nodeDy)}{\pgfxy(\nodeEx,\nodeEy)}{\textedgeoffsetright}{\pgfbox[center,center]{$(s+4r)+5(r+4r_a)$}}
\pgfputlabelrotated{.54}{\pgfxy(\nodeDx,\nodeDy)}{\pgfxy(\nodeEx,\nodeEy)}{\textedgeoffsetrightdown}{\pgfbox[center,center]{$\left(=(s+5r)+4(r+5r_a)\right)$}}
\pgfputlabelrotated{.56}{\pgfxy(\nodeDx,\nodeDy)}{\pgfxy(\nodeFx,\nodeFy)}{\textedgeoffset}{\pgfbox[center,center]{$(s+4r)+6(r+4r_a)$}}
\pgfputlabelrotated{.56}{\pgfxy(\nodeDx,\nodeDy)}{\pgfxy(\nodeFx,\nodeFy)}{\textedgeoffsetdown}{\pgfbox[center,center]{$\left(=(s+6r)+4(r+6r_a)\right)$}}
\usebox{\mynetworkC}
\end{pgfpicture}\vspace{\vspacingBetnGraphs}

\begin{pgfpicture}{0cm}{0cm}{\needWidth}{\needHeight}
\pgfputlabelrotated{.51}{\pgfxy(\nodeEx,\nodeEy)}{\pgfxy(\nodeFx,\nodeFy)}{\textedgeoffsetleft}{\pgfbox[center,center]{$(s+5r)+6(r+5r_a)$}}
\pgfputlabelrotated{.51}{\pgfxy(\nodeEx,\nodeEy)}{\pgfxy(\nodeFx,\nodeFy)}{\textedgeoffsetleftdown}{\pgfbox[center,center]{$\left(=(s+6r)+5(r+6r_a)\right)$}}
\usebox{\mynetworkC}
\end{pgfpicture}
\vspace{-2in}
\label{fig:toy_c}
\end{minipage}
}}
\end{minipage}
\caption{Secret sharing across the network of Fig.~\ref{fig:toy_a}, for $n=6$ and $k=2$: (a) the \csota, employing separate secure message transmissions from the dealer to each participant, and (b) our new \sneaka. The text on an edge is the data passed by the node  with the lower index to the node with the higher index. See Example~\ref{ex:toy} for more details.
}
\label{fig:toy_our}
\end{figure*}
Several cryptographic protocols in the literature require execution of one or more instances of secret sharing among all the participants. These include protocols for secure multiparty-computation~\cite{ben1988completeness,chaum1988multiparty,cramer1996multi,damgaard2010perfectly}, secure key management~\cite{pedersen1991threshold,marsh2004codex}, general Byzantine agreement between all participants~\cite{rabin1983randomized,ben1988completeness,feldman1988optimal,ingemarsson1990protocol},  proactive secret sharing~\cite{ostrovsky1991withstand,herzberg1995proactive}, and secure archival storage~\cite{storer2009potshards}. For instance, under the celebrated Ben-Or-Goldwasser-Wigderson (BGW) protocol ~\cite{ben1988completeness} for secure-multiparty function computation, the initialization step requires $n$ instances of secret sharing with all participants, and every multiplication operation requires $2n$ additional instances.

Most protocols including those listed above assume that the dealer has \textit{direct} secure communication links to every participant. In this case, the dealer can compute the shares as per Shamir's scheme~\cite{shamir1979share} and directly pass the shares to the respective participants. Such a setting is depicted in Fig.~\ref{fig:fully_connected} for the parameters $(n=6,\,k=2)$. In many situations, however, the dealer may not have direct communication links with every participant; instead, the dealer and the participants may form a general network. Fig.~\ref{fig:toy_a} depicts such a scenario. The network is described by a graph $\mathcal{G}$ with $(n+1)$ nodes. These $(n+1)$ nodes comprise the dealer and the $n$ participants. An edge in this graph implies a communication link between its two end-points, while the absence of an edge denotes the non-existence of any direct communication link. We make the standard assumption that the communication links between the participants are secure. We will say that a participant is `directly connected to the dealer' if there exists an edge from the dealer to that participant. 

Under a general network $\mathcal{G}$, all communication between the dealer and any participant who is not directly connected to it, must pass through other participants in the network. This poses the challenge of not leaking any additional information to any participant while disseminating the shares over the network. 

Existing methods use separate secure message transmissions (SMT) from the dealer to each participant across the network~\cite{dolev1993perfectly}.  Under such a solution, in order to communicate the designated share to a participant, the dealer treats this share as a secret, employs Shamir's scheme to compute $k$ shares of this secret, and communicates these $k$ shares to the participant through $k$ node-disjoint paths. (This solution is described in more detail in Section~\ref{sec:literature_naive}.) Thus solution requires a significant coordination in the network in setting up the node-disjoint paths from the dealer to every participant. It also incurs a high communication cost, since the dealer needs to transmit shares across the network separately for every participant. 

In this paper, we present a distributed and communication-efficient algorithm for secret share dissemination across a network, which we call \sneaka\footnote{`SNEAK' standing for `Secret-sharing over a Network with Efficient communication And distributed Knowledge-of-topology'.}. We analyze the performance of \sneak and compare it to the state-of-the-art, i.e., the \csotas. In addition to being distributed, \sneak provides significant gains in terms of the communication cost and the amount of randomness required. 
On the other hand, while the \csotas requires the graph to satisfy a certain ``$k$-connected-dealer'' condition (which is in fact necessary for the feasibility of secret sharing), \sneak imposes a stronger condition on the network, which we call the $k$-propagating-dealer condition, that will be formalized in the sequel.

We now present a toy example illustrating the existing \csota and \sneaka proposed in this paper.
\begin{example}\label{ex:toy}
{\it
Consider the network depicted in Fig.~\ref{fig:toy_a}. Let $n=6$ and $k=2$, with the alphabet of operation as the finite field $\mathbb{F}_7$. Under Shamir's scheme of encoding the secret $s$, the share $t_i~(1\leq i \leq 6)$ for participant $i$ is \[t_i = s + ir~,\] where $r$ is a value chosen by the dealer uniformly at random from the alphabet $\mathbb{F}_7$. While the dealer can directly pass the shares $t_1$ and $t_2$ to participants $1$ and $2$ respectively, the difficulty arises in communicating shares to the remaining participants with whom the dealer does not have direct communication links. For instance, if the dealer tries to pass share $t_3$ to participant $3$ by simply communicating $t_3$ along the path `dealer $\rightarrow$ $1$ $\rightarrow$ $3$', then participant $1$ gains access to two shares, $t_1$ and $t_3$. Using these two shares, participant $1$ can recover the secret $s$, thus violating the $(k-1)$-collusion resistance requirement.

The \csota is illustrated in the sequence of steps depicted in Fig.~\ref{fig:toy_b}. In order to pass the share $t_3$ to participant $3$, the dealer chooses another random value $r_3$, passes $(t_3 + r_3)$ along the path `dealer $\rightarrow$ $1$ $\rightarrow$ $3$', and $r_3$ along the path `dealer $\rightarrow$ $2$ $\rightarrow$ $3$'. Now, participant $3$ can recover $t_3$, and no participant gains any additional information about the secret $s$ in this process. In a similar manner, the dealer can communicate $t_i~(4 \leq i \leq 6)$ to participant $i$ through $k\!=\!2$ node-disjoint paths as shown in the figure. 

Observe that the \csota transmits data across several hops in the network in every step, but the data transmitted in any step is never used in subsequent steps of the protocol. Thus, in order to design an efficient and distributed algorithm, one may wish to propagate data in a manner that allows its subsequent reuse downstream. This is the key idea underlying \sneaka proposed in this paper, which is illustrated in the sequence of steps in Fig.~\ref{fig:toy_c}. Here, the dealer first draws two values $r$ and $r_a$ uniformly at random from $\mathbb{F}_7$. The dealer then passes the two values $(s+r)$ and $(r+r_a)$ to node $1$, and the two values $(s+2r)$ and $(r+2r_a)$ to node $2$. Upon receiving its data, each node passes a particular linear combination of its received data to each of its downstream neighbours. For instance, node $1$ passes $(s+r)+3(r+r_a)$ to node $3$, which can equivalently be written as $(s+3r)+(r+3r_a)$. Node $2$ passes $(s+2r)+j(r+2r_a)~(=(s+jr)+2(r+jr_a))$ to node $j \in \{3,4\}$ respectively. Node $3$ can thus recover the two values $(s+3r)$ and $(r+3r_a)$ from the data it receives. Similarly, as shown in the sequence of steps depicted in Fig.~\ref{fig:toy_c}, every node $i \in \{1,\ldots,6\}$ can recover its requisite share $(s+ir)$, along with a random counterpart $(r+ir_a)$ which is used to disseminate shares further downstream.
Note that in Fig.~\ref{fig:toy_c}, the expression written above an edge is the linear combination that is transmitted, and the corresponding expression written in the parenthesis below that edge is a simple rewriting of the data transmitted. 

We can see that \sneaka is completely distributed requiring knowledge of only one-hop neighbours as opposed to the \csota which requires the knowledge of the global topology in order to set-up communication over node-disjoint paths. Also, \sneaka  requires a communication of only $12$ values, as opposed to $24$ under the \csota. The number of random values generated under \sneaka is only $2$,  whereas the \csotas requires generation of $5$ random values.}
\end{example}

The remainder of the paper is organized as follows. Section~\ref{sec:model} provides a formal description of the system model and summarizes the results of this paper. Section~\ref{sec:literature} reviews related literature. Section~\ref{sec:algorithm} describes \sneaka. Section~\ref{sec:analysis} presents a comparative analysis of the \csota, \sneaka, and the lower bounds in terms of the communication-cost and the randomness requirements. Section~\ref{sec:conclusion} presents conclusions and discusses open problems. 

\section{System Model and Summary of Results}\label{sec:model}
\subsection{Secret Sharing in a General Network}
The dealer possesses a secret $s$ that is drawn from some alphabet $\mathcal{A}$, and wishes to pass shares of this secret to $n$ participants. The dealer and the participants form a communication network, denoted by graph $\mathcal{G}$. The graph $\mathcal{G}$ has $(n+1)$ nodes comprising the dealer and the $n$ participants, and an edge in the graph denotes a secure and private communication link between the two end-points. Thus, at times, we will also refer to a participant as a node of the graph. We will also use the terms `network' and `graph' interchangeably, and `link' and `edge' interchangeably. The problem is to design a protocol which will allow the dealer to pass shares (of the secret) to the $n$ participants, meeting the requirements of \textit{$(k-1)$-collusion-resistance} and \textit{$k$-secret-recovery} (described in Section~\ref{sec:intro}). All the participants are assumed to be honest-but-curious, i.e., they follow the protocol correctly, but may store any accessible data in order to gain information about the secret. The edges in the graph $\mathcal{G}$ are allowed to be directed or undirected: a directed edge implies existence of only a one way communication link, while an undirected edge implies direct communication links both ways. The parameters $n$ and $k$ are assumed to satisfy $n \geq k >1$, since $n\leq k-1$ prohibits the secret from ever being recovered, while $k=1$ degenerates the problem into a trivial case wherein no secrecy is required.

We now discuss a condition that the graph $\mathcal{G}$ must necessarily satisfy for \textit{any} algorithm to successfully perform secret sharing on it, which directly follows from~\cite{dolev1993perfectly}.
\begin{definition}[$m$-connected-dealer]
A graph with $(n+1)$ nodes (the dealer and $n$ participants) satisfies the $m$-connected-dealer condition for a positive integer $m$ if each of the $n$ participants in the graph either has an incoming edge directly from the dealer or has at least $m$ node-disjoint paths from the dealer to itself.
\end{definition}

\begin{prop}[Necessary condition~\cite{dolev1993perfectly}]\label{prop:necessary}
For any graph $\mathcal{G}$, a necessary condition for any algorithm to perform $(n,k)$ secret sharing is that $\mathcal{G}$ satisfies the $k$-connected-dealer condition.
\end{prop}

Thus no algorithm can operate successfully on all network topologies, and must require the graph $\mathcal{G}$ to obey at least the $k$-connected-dealer condition. 


\subsection{Class of Networks Considered} \label{sec:model_class_of_networks}
We saw above in Proposition~\ref{prop:necessary} that feasibility of secret sharing on a graph $\mathcal{G}$ requires $\mathcal{G}$ to satisfy the $k$-connected-dealer condition. \sneaka requires the communication network $\mathcal{G}$ to satisfy a stronger condition, which we term the $k$-propagating-dealer condition.

\begin{definition}[$m$-propagating-dealer]
A graph with $(n+1)$ nodes (the dealer and $n$ participants) satisfies the $m$-propagating-dealer condition for a positive integer $m$ if there exists an ordering of the $n$ participants in the graph such that every node either has an incoming edge directly from the dealer or has incoming edges from at least $m$ nodes preceding it in the ordering.
\end{definition}

\begin{figure}[t!]
\centering
\subfloat[Layered network]{
\includegraphics[width=.28\textwidth]{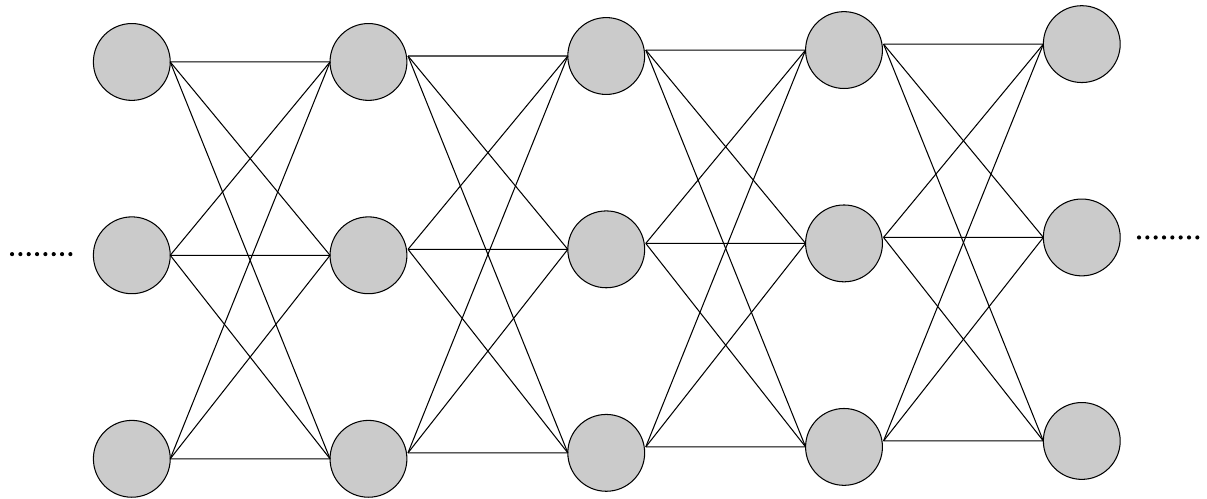}
\label{fig:newtorks_propagating_dealer_layered}
}~~
\subfloat[Backbone network]{
\includegraphics[width=.18\textwidth]{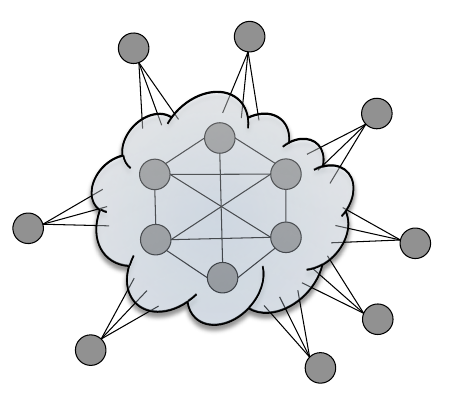}
\label{fig:newtorks_propagating_dealer_backbone}
}~~
\subfloat[One-dimensional geometric network (e.g., sensors deployed on a border)]{
\includegraphics[width=.45\textwidth]{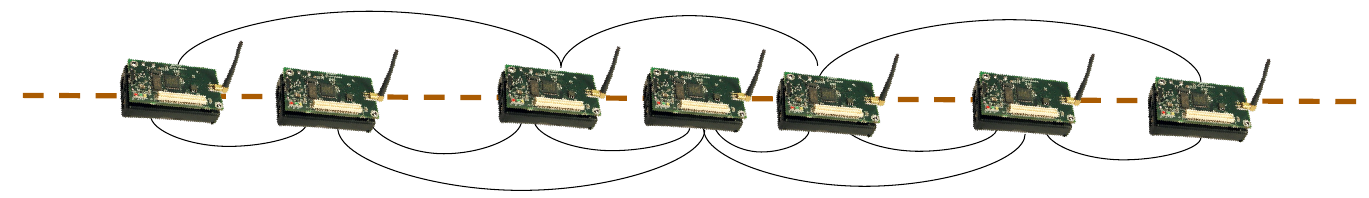}
\label{fig:newtorks_propagating_dealer_geometric}
}
\caption{Examples of undirected networks satisfying the $3$-propagating-dealer condition. Any node in the network may be the dealer.}
\label{fig:networks_propagating_dealer}
\end{figure}

As an illustration of the $m$-propagating-dealer condition, consider the network of Example~\ref{ex:toy} (Fig.~\ref{fig:toy_a}). This network satisfies the $2$-propagating-dealer condition, with the ordering $1,\,2,\,3,\,4,\,5,\,6$ (observe that this is also the order in which the participants receive their shares under \sneaka as shown in Fig.~\ref{fig:toy_c}). 

We enumerate below examples of a few classes of graphs that satisfy the $m$-propagating dealer condition. 

Fig.~\ref{fig:networks_propagating_dealer} depicts three examples of graphs that satisfy the $3$-propagating dealer condition. These examples can be generalized to the following classes of graphs. The first three classes consider undirected graphs and permit any of the nodes to be the dealer.

 (a) Layered networks (Fig.~\ref{fig:newtorks_propagating_dealer_layered}): Each layer contains at least $m$ nodes, and each node is connected to all nodes in the neighbouring layers. An ordering that satisfies the $m$-propagating-dealer condition is the ordering of the nodes with respect to the distance (in terms of number of hops) from the dealer.
 
 (b) Networks with a backbone (Fig.~\ref{fig:newtorks_propagating_dealer_backbone}): A subset of nodes, termed the `backbone', form a densely connected subgraph (that satisfies the $m$-propagating-dealer condition), and every node outside the backbone is connected directly to at least $m$ nodes in the backbone. An ordering that satisfies the $m$-propagating-dealer condition is: the ordering under the backbone subgraph, followed by all remaining nodes not in the backbone in any order.
 
 (c) $m$-connected one-dimensional geometric networks (Fig.~\ref{fig:newtorks_propagating_dealer_geometric}): A one-dimensional geometric network is formed by arranging the nodes (in an arbitrary fashion) along a line, and connecting a pair of nodes by an edge if the distance (number of hops) between is smaller than a fixed threshold. A one-dimensional geometric network that satisfies the $m$-connected-dealer condition also satisfies the $m$-propagating dealer condition. In this case, an ordering that satisfies the $m$-propagating-dealer condition is the arrangement of the nodes in an ascending order of their euclidean distance from the dealer.
 
 (d) Directed acyclic graphs: Any directed acyclic graph that satisfies the $m$-connected-dealer condition also satisfies the $m$-propagating-dealer condition. The root must be the dealer, and any topological ordering of the DAG suffices as the requisite node-ordering. E.g., the directed version of the graph of Fig.~\ref{fig:toy_a}, with the edges directed from the left to the right, falls in this class.


We note that while the necessity of the $k$-propagating-dealer condition under \sneaka requires the \textit{existence} of some such ordering of the nodes, the execution of the algorithm is completely distributed and oblivious to the actual ordering. We also note that while \sneak requires the graph to satisfy the $k$-propagating dealer condition, it is \textit{robust to the network topology}, i.e., it satisfies the $(k-1)$-collusion-resistance property over any arbitrary network topology. This robustness property is important since the secrecy of the secret is paramount, and in practice, the structure of the network may not be known beforehand. Moreover, under a dynamic network, the graph structure may also vary with time (even during the execution of the algorithm), thus further motivating robustness. 

\comment{
To prove:
a) Random geometric graphs~\cite[Section 2.6 and Theorem 2.16]{wormald1999models}.
b) Erdos-Renyi graphs conditioned on k-connectivity
c) http://en.wikipedia.org/wiki/Barab
d) some kind of a hierarchical network (?)
e) expander graph
On the other hand, an example of a graph that satisfies the $k$-connected-dealer condition but not the $k$-propagating-dealer condition is a cycle (with $k=2$).
}

Apart from the parameters $n$ and $k$, an \textit{additional parameter} ``$d$'' is associated to \sneaka. We saw earlier that the $k$-connected-dealer condition is necessary for \textit{any} secret sharing algorithm, and \sneaka requires the $k$-propagating-dealer condition to be satisfied. Now, assuming that these necessary conditions have been met, one would intuitively expect the efficiency of the algorithm to be higher if the graph has a higher connectivity. The parameter $d$ captures this intuition: \sneaka takes the parameter $d~(\geq k)$ as input, and under the assumption that the graph satisfies the \textit{$d$-propagating-dealer} condition, achieves a greater communication efficiency.


\subsection{Summary of Results}
This paper presents a distributed algorithm, called \sneak, that takes parameters $n$, $k$ and $d~(\geq k)$ as input, and enables a dealer to disseminate shares of a secret to $n$ participants forming a general network $\mathcal{G}$, such that the properties of
(i) $k$-secret-recovery (when $\mathcal{G}$ satisfies the $d$-propagating-dealer condition), and (ii) $(k-1)$-collusion-resistance (irrespective of the network topology) are satisfied. The algorithm is completely distributed and each node needs to know only the identities of its neighbours, and it is efficient with respect to the communication cost and the amount of randomness required.

For any $(n,k)$ and any graph $\mathcal{G}$ with $(n+1)$ nodes, we derive and compare (a) information-theoretic lower bounds on the total communication cost under {any} algorithm, (b) lower bounds on the communication cost under the \csota, and (c) communication cost under \sneaka.
Using these results, we establish the communication efficiency of \sneaka. \sneaka has particularly useful implications on bounded degree graphs.\footnote{Bounded degree graphs have their maximum degree upper bounded by a constant that is independent of $n$.} 
For networks with bounded degree and satisfying the $k$-propagating-dealer condition, the communication cost of \sneaka is  within a constant (multiplicative) factor of the information-theoretic lower bound, and is $\Theta(n)$ in the worst-case. In contrast, the communication-cost of the \csotas is lower bounded as $\Omega(n \log n)$, and is $\Omega(n^2)$ in the worst case. Moreover, the amount of randomness required under \sneaka is $\Theta(1)$, while that required under the \csota is $\Theta(n)$. 


Appendix~\ref{app:extensions} presents three heuristic techniques to extend \sneaka algorithm to networks where the $k$-propagating-dealer condition is not satisfied. Appendix~\ref{app:extensions} also contains an extension of the algorithm to handle active adversaries, and an extension to support two-threshold secret sharing.

\subsection{Notational Conventions}
A vector will be treated as a column vector by default, and a row vector will be written as the transpose of the corresponding column vector. The transpose of a vector or a matrix will be denoted by a superscript $T$. For any integer $\ell \geq 1$, $[\ell]$ will represent the set $\{1,\ldots,\ell\}$. For any participant $j~(1\leq j \leq n)$, the set of its neighbours will be denoted by $\mathcal{N}(j)$. In case of a directed graph, $\mathcal{N}(j)$ will denote the set of nodes to which node $j$ has an \text{outgoing} edge. The dealer will be denoted as $D$, and the set of neighbours of the dealer by $\mathcal{N}(D)$. We will say that a node $j$ is directly connected to the dealer if $j \in \mathcal{N}(D)$.

\section{Related Literature}\label{sec:literature}
\subsection{Shamir's Secret Sharing Protocol}\label{sec:literature_shamir}
We first give a brief review of Shamir's secret sharing protocol~\cite{shamir1979share}. To this end, we assume that the dealer has a direct (secure) communication link with every participant (as in the example in Fig.~\ref{fig:fully_connected}).

Assume that the secret $s$ is drawn from some finite field $\mathbb{F}_q$ of size $q~(>n)$. The dealer chooses $(k-1)$ values $\lbrace r_i \rbrace_{i=1}^{k-1}$ independently and uniformly at random from $\mathbb{F}_q$. Define a $k$-length vector $\boldsymbol{m}$ as\,\footnote{To suit the description of the algorithm developed subsequently in this paper, we present a matrix-based representation for Shamir's protocol instead of the customary polynomial based representation. We will remark upon the connection between the two representations in the end.}
\beq \boldsymbol{m}^T = \left[ s ~ r_1 ~ r_2 ~ \cdots ~ r_{k-1}\right]~.\eeq 
Next, define a set of $n$ vectors $\lbrace \boldsymbol{\psi}_i \rbrace_{i=1}^{n}$, each of length $k$, as \beq \boldsymbol{\psi}_i^T = [1~i~i^2~\cdots~i^{k-1}]~. \eeq 
The share $t_i$ of participant $i$ is simply the inner product \beq t_i = \boldsymbol{\psi}_i^T \boldsymbol{m}~.\label{eq:shamir_original}\eeq It can be verified that the vectors $\{\boldsymbol{\psi}_i\}_{i \in \mathcal{I}}$ are all linearly independent. It follows that for any set $\mathcal{I} \subseteq [n]$ of cardinality $k$, the secret $s$ can be recovered from the set of values $\lbrace \boldsymbol{\psi}_i^T \boldsymbol{m} \rbrace_{i \in \mathcal{I}}$. Furthermore, it can also be verified that for any set $\mathcal{I}' \subseteq [n]$ of cardinality smaller than $k$, the set $\lbrace \boldsymbol{\psi}_i^T \boldsymbol{m} \rbrace_{i \in \mathcal{I}'}$ provides no knowledge about $s$.

Under the assumption that the dealer has direct communication links with each of the $n$ participants, the dealer can simply pass $t_i$ to participant $i$ ($\in [n]$). 

\begin{remark}
To see the Shamir's secret sharing scheme in the conventional polynomial representation~\cite{shamir1979share}, note that each share $t_i$ can be seen as an evaluation of the $(k-1)$ degree polynomial with the secret $s$ as its constant term and the remaining $(k-1)$ coefficients chosen uniformly at random and independent of the secret. Thus, $k$ evaluations are necessary and sufficient to recover the polynomial and hence the secret. This provides the $k$-secret-recovery and $(k-1)$-collusion-resistance properties. 
\end{remark}

This completes the description of Shamir's secret sharing protocol. There are numerous other extensions and generalizations of Shamir's secret sharing protocol in the literature, and the reader is referred to~\cite{stinson1992explication,beimel2011secret} and references therein for more details. We now describe the \csota that addresses the situation when the dealer and participants form a general communication network.

\subsection{The \csota}\label{sec:literature_naive}
This section describes a scheme for secret sharing over a general communication network employing secure message transmissions from dealer to each participant~\cite{dolev1993perfectly}. Fig.~\ref{fig:toy_b} in Example~\ref{ex:toy} is an example of such a solution. Under this solution, the dealer first encodes the secret $s$ into $n$ shares $\lbrace t_\ell \rbrace_{\ell=1}^{n}$ using Shamir's secret sharing scheme~\eqref{eq:shamir_original}. The size $q$ of the underlying finite field $\mathbb{F}_q$ is assumed to be greater than $n$. To every node $\ell$ directly connected to the dealer, the dealer directly passes its share  $t_\ell$. To disseminate shares to the remaining nodes, the dealer performs the following actions, once separately for each remaining node. Let $\ell$ now denote a node that is not connected directly to the dealer. The dealer applies Shamir's secret sharing scheme treating $t_\ell$ as a secret, and computes $k$ shares $\lbrace u_{\ell,j} \rbrace_{j=1}^{k}$, as 
\beq u_{\ell,j} = [1~j~j^2~\cdots~j^{k-1}] \left[t_{\ell}~r_{{\ell},1}~r_{{\ell},2}~\cdots~r_{{\ell},k-1}\right]^T~, \eeq 
where the values $\lbrace r_{{\ell},1},\ldots,r_{{\ell},k-1} \rbrace$ are chosen independently and uniformly at random from $\mathbb{F}_q$. The dealer then finds $k$ node-disjoint paths (from itself) to node $\ell$, and passes $u_{\ell,j}$ along the $j^\text{th}$ path $(1\leq j \leq k)$. At the end of these transmissions, node $\ell$ receives $\lbrace u_{\ell,j} \rbrace_{j=1}^{k}$ from which it can recover its share $t_\ell$. Moreover, since each of the random values are independent, no participant can obtain any information about any other participant's share, or any additional information about the secret $s$. This process is repeated once for every node that is not connected directly to the dealer.


The solution described above requires transmission of data across $k$ node-disjoint paths \textit{once separately for every node} that is not connected directly to the dealer. Thus this solution is not distributed, and furthermore is not efficient in terms of communication and the randomness cost. 

We note that the communication efficiency of this solution can be improved if more than $k$ node-disjoint paths are available, by employing \textit{two-threshold secret sharing}~\cite{franklin1992communication} over these node-disjoint paths. Under this setting, for any given participant $i$, let us suppose there are $w_i~(\geq k)$ node-disjoint paths from the dealer to node $i$. The dealer chooses a value $w~(\in \{k,\ldots,w_i\})$, encodes the share of participant $i$ into $w$ chunks in a manner~\cite{franklin1992communication} that satisfies $w$-secret-recovery and $(k-1)$-collusion-resistance, and passes these chunks via the $w$ shortest node-disjoint paths to participant $i$. The dealer chooses $w$ such that the amount of communication in transmitting the share to participant $i$ is minimized; the special case of choosing $w=k$ for all participants is equivalent to the procedure described in the previous paragraphs. The analysis and comparisons performed subsequently in Section~\ref{sec:analysis} will consider this more efficient two-threshold version of the \csotas.

\subsection{Network Coding and Distributed Storage}\label{sec:literature_nc}
The problem of secret sharing over a general communication network can also be cast as a specific instance of a \textit{network coding} problem~\cite{ahlswede2000network}, requiring security from eavesdropping on the nodes. This casting can be performed in the following manner. The dealer is the source node, and the secret $s$ is the message. The network graph in the network coding problem is identical to that in the secret sharing problem, but with a set of ${n \choose k}$ additional nodes that act as the sinks. Each of the ${n \choose k}$ sinks is connected to a distinct subset of $k$ participants, and has one directed link of infinite capacity coming in from each of the corresponding $k$ participants. Each sink must recover the entire message: this requirement corresponds to the condition of $k$-secret-recovery. To satisfy the $(k-1)$-collusion-resistance property, a compromise of upto $(k-1)$ arbitrary nodes (excluding the source and the sinks) to a passive eavesdropper should reveal no information about the message. In this manner, the secret sharing problem is equivalent to a network coding problem requiring secrecy from an eavesdropper that can gain access to a subset of the nodes. However, with respect to this setting, very little appears to be known in the network coding literature.

To the best of our knowledge, the literature on secure network coding (e.g.,~\cite{yao2010network,ngai2009secure,cai2011secure,feldman2004capacity}) considers the setting where the eavesdropper gains access to a subset of the \textit{links}. The problem of node compromise is typically treated as a case of link compromise by allowing the eavesdropper to gain access to all links that are incident upon the compromised nodes. In~\cite{cai2011secure,feldman2004capacity}, authors consider the setting wherein a collection of subsets of the links is specified, and an eavesdropper may gain access to precisely one of these subsets. However, the scheme requires the knowledge of the network topology, and is computationally expensive. Moreover, 
the scheme requires the graph to satisfy a particular condition, which is almost always violated in our problem setting. Communication-efficient algorithms to secure a network from an eavesdropper having access to a \textit{bounded} number of links are provided in~\cite{yao2010network,ngai2009secure}. Given the network topology, the actions to be performed at the nodes can be derived in a computationally efficient manner. However, these algorithms communicate a message of size equal to the difference between the largest message that can be sent in the absence of secrecy requirements, and the bound on number of compromised links. Under our problem setting, this difference is generally zero or smaller (e.g., the difference is $-2$ in the network of Fig.~\ref{fig:toy_a}), thus rendering these algorithms inapplicable.

The \sneaka algorithm constructed in the present paper thus turns out to be an instance of a secure network coding problem that admits an explicit solution that is distributed, communication efficient, and provides deterministic (probability $1$) guarantees. Furthermore, the solution handles the case of nodal eavesdropping, about which very little appears to be known in the literature.


\sneaka is based on a variant of the \textit{Product-Matrix codes}~\cite{ourProductMatrix} which were originally constructed for distributed storage systems. These codes possess useful properties that \sneaka exploits in the present context. The product-matrix codes are a practical realization of the concept of `Regenerating codes'~\cite{YunDimKanJournal} proposed for distributed storage. To date, apart from the MDS codes of~\cite{ourAllerton}, these are the only known constructions of regenerating codes that are scalable (i.e., other parameters of the system impose no constraints on the total number of nodes in the system). It turns out that this scalability property is an essential ingredient for our problem. Secure versions of the product-matrix codes were constructed in~\cite{ourGlobecom2011,ourISIT2012}. The reader familiar with the literature on regenerating codes for distributed storage may recognize later in the paper that we employ the minimum-bandwidth (MBR) version, and not the minimum-storage (MSR) version, of the product-matrix codes~\cite{ourProductMatrix}. We make this choice to guarantee secrecy from honest-but-curious participants, who may store all the data that they receive, a characteristic of the MBR point on the storage-bandwidth tradeoff~\cite{YunDimKanJournal}.

\section{Algorithm for Secret Sharing Over General Networks}\label{sec:algorithm}
This section presents the main result of the paper. Consider a network $\mathcal{G}$ that obeys the \textit{$d$-propagating-dealer condition} for some parameter $d~(\geq k)$. The secret $s$ belongs to the alphabet $\mathcal{A}$, and we assume that $\mathcal{A} = \mathbb{F}_q^{d-k+1}$, for any arbitrary field size $q>n$. Thus we can equivalently denote the secret as a vector $\mathbf{s} = [s_1~s_2~\cdots~s_{d-k+1}]^T$ with each element of this vector belonging to the finite field $\mathbb{F}_q$.

\subsection{Initial Setting up by the Dealer}
The dealer first constructs an $(n \times d)$ Vandermonde matrix $\Psi$, with the $i^{th}~(1\leq i \leq n)$ row of $\Psi$ being 
\beq \boldsymbol{\psi}_i= [1~i~i^2~\cdots~i^{d-1}]^T~.\eeq
The vector $\boldsymbol{\psi}_i$ is termed the \textit{encoding vector} of node $i$.

\def\Msa{s_A}
\def\Mrb{\boldsymbol{r_a}}
\def\Msc{\boldsymbol{s_B}}
\def\Mrd{R_b}
\def\Mre{R_c}

Next, the dealer constructs a $(d \times d)$ \textit{symmetric} matrix $M$ comprising the secret $\mathbf{s}$ and a collection of randomly generated values as follows:
\def\Mstruct{\left[\begin{array}{ccc} \Msa & \Mrb^T & \Msc^T \\ \Mrb & \Mrd & \Mre^T \\\Msc & \Mre & 0\end{array}\right]}

\bea M &=& \Mstruct \label{eq:M} \\
&& ~~\underbrace{\!\!\!\!\!}_{1}\,\underbrace{~~~~~~}_{k-1}\,\underbrace{~~~~~~~~}_{d-k}  \nonumber\\
&& ~~\underbrace{~~~~~~~~~~~~~~~~~~~~~}_{d} \nonumber
\eea
 where the depicted sub-matrices of $M$ are\\
 \begin{itemize}
\item $\Msa = s_{d-k+1}$ is a scalar,
\item $\Msc = [ s_1 \cdots s_{d-k} ]^T$ is a vector of length $(d-k)$,
\item $\Mrb$ is a vector of length $(k-1)$ with its entries populated by random values,
\item $\Mrd$ is a $((k-1) \times (k-1))$ \textit{symmetric} matrix with its $\frac{k(k-1)}{2}$ distinct entries populated by random values,
\item $\Mre$ is a $((d-k) \times (k-1))$ matrix with its $(k-1)(d-k)$ entries populated by random values.
\end{itemize}
These random values are all picked independently and uniformly from $\mathbb{F}_q$. Note that the total number of random values $R$ in matrix $M$ is
\bea R &=& (k-1) +\frac{k(k-1)}{2} + (k-1)(d-k)\nonumber\\&=&  (k-1)d-{k-1\choose 2}~\label{eq:num_random}.\eea
The entire secret is contained in the components $\Msa$ and $\Msc$ as $\mathbf{s}^T = [s_1 \cdots s_{d-k+1}] = [ \Msc^T~~\Msa ]$.

Observe that the structure of $M$ as described in~\eqref{eq:M}, along with the symmetry of matrix $\Mrd$, makes the matrix $M$ \textit{symmetric}.

The share $\mathbf{t}_j$ for participant $j~(1\leq j \leq n)$ is a vector of length $(d-k+1)$:
\beq \boldsymbol{t}_j^T = \boldsymbol{\psi}_j^T \left[\begin{array}{cc}\Msa&\Msc^T\\\Mrb&\Mre^T \\\Msc&0~\end{array}\right]~. \label{eq:share}\eeq  We will show subsequently in Theorem~\ref{thm:recovery} that any $k$ of these shares suffice to recover the entire secret.

\begin{remark}
To see these shares in the conventional polynomial representation of Shamir's secret sharing scheme, recall that the vector $\boldsymbol{\psi}_j^T$ is drawn from a Vandermonde matrix. Thus each entry of $\boldsymbol{t}_j$ in~\eqref{eq:share} can be seen as the evaluation of a polynomial at value $j$. Thus there is one polynomial for each secret value $s_i~(1 \leq i \leq d-k+1)$, with the corresponding secret symbol as its constant term and the remaining coefficients independent of the secret value $s_i$.
\end{remark}

\begin{example}\label{ex:mainalgo_init}
\it
Consider the setting of Example~\ref{ex:toy} (Fig.~\ref{fig:toy_c}) where $n=6,~k=2,~d=2$. Here \[
M = 
\left[\begin{array}{cc}
s&r\\
r&r_a
\end{array}\right],
\]
and for every $j~(1\leq j \leq n)$, $\boldsymbol{\psi}_j^T = [1~~j]$ and the share for participant $j$ is $\boldsymbol{t}_j^T=[s+jr]$.
\end{example}

\subsection{Communication across the Network}
Algorithm~\ref{algo:comm} describes the communication protocol to securely transmit the shares $\lbrace \boldsymbol{t}_j \rbrace_{j=1}^{n}$ to the $n$ participants.

\begin{algorithm}[H]
\begin{itemize}
\settowidth{\itemindent}{\bf Dealer:}
\item[\bf Dealer:] For every $j \in \mathcal{N}(D)$, compute and pass the $d$-length vector $\boldsymbol{\psi}_j^T M$ to participant $j$.
\vspace{.3cm}
\settowidth{\itemindent}{\bf Participant $\boldsymbol{i \in \mathcal{N}(D)}$:}
\item[\bf Participant $\boldsymbol{\ell \in \mathcal{N}(D)}$:]
Wait until receipt of data $\boldsymbol{\psi}_{\ell}^T M$ from the dealer. Upon receipt, perform the following actions. For every $j \in \mathcal{N}(\ell)$, compute the inner product of the data $\boldsymbol{\psi}_{\ell}^T M$ with the encoding vector $\boldsymbol{\psi}_j$ of participant $j$. Transmit the resulting value $\boldsymbol{\psi}_{\ell}^T M \boldsymbol{\psi}_j$ to participant $j$.
\vspace{.3cm}
\item[\bf Participant $\boldsymbol{\ell \notin \mathcal{N}(D)}$:] Wait until receipt of one value each from any $d$ neighbours, and then perform the following actions (if more than $d$ neighbours pass data, retain data  from some arbitrary $d$ of these nodes). Denote this set of $d$ neighbours as $\lbrace i_1,\ldots, i_d\rbrace$, and the values received from them as $\lbrace \sigma_1,\ldots,\sigma_d\rbrace$ respectively. Compute the vector \beq \boldsymbol{v} = \left[ \begin{array}{c} ~~~~~~\boldsymbol{\psi}_{i_1}^T~~~~~~ \\\vdots\\\boldsymbol{\psi}_{i_d}^T\end{array}\right]^{-1} \left[\begin{array}{c} \sigma_1 \\ \vdots \\ \sigma_d \end{array}\right].\nonumber\eeq
For every neighbour $i \in \mathcal{N}(\ell)$ from whom you did not receive data, compute and pass the inner product $\boldsymbol{v}^T \boldsymbol{\psi}_i$ to participant $i$.
\end{itemize}
\caption{Communication Protocol}\label{algo:comm}
\end{algorithm}

\begin{remark}
In order to reduce the communication cost, one would like to ensure that a participant receives data from no more than $d$ of its neighbours. This can be ensured via a simple handshaking protocol between neighbours, wherein a participant who is ready to transmit data to its neighbours, queries the neighbours for the requirement of the respective transmissions, prior to actually sending the data.
\end{remark}

\begin{example}
\it
Consider the setting of Example~\ref{ex:toy} (Fig.~\ref{fig:toy_c}), wherein $n=6,~k=2,~d=2$. The values of $M$, $\boldsymbol{\psi}_j$ and $\boldsymbol{t}_j$~$(1\leq j \leq n)$ under this setting are specified in Example~\ref{ex:mainalgo_init}. For the given network, we have $\mathcal{N}(D)=\{1,2\}$. As per Algorithm~\ref{algo:comm}, participant $j\in\{1,2\}$ receives $\boldsymbol{\psi}_j^TM=[s+jr~~~r+jr_a]$ directly from the dealer. Now let us focus on participant $3$. Since participant $3$ is a neighbour to participants $1$ and $2$, following Algorithm~\ref{algo:comm}, participant $j\in\{1,2\}$ passes $\boldsymbol{\psi}_j^TM\boldsymbol{\psi}_3 = (s+jr)+3(r+jr_a)$ to participant $3$. Participant $3$ thus receives the two values $\sigma_1=(s+r)+3(r+r_a)$ and $\sigma_2=(s+2r)+3(r+2r_a)$ from neighbours $i_1=1$ and $i_2=2$. Using the fact that $\boldsymbol{\psi}_{1}^T=[1~~1]$ and $\boldsymbol{\psi}_{2}^T=[1~~2]$, it computes 
\[
\boldsymbol{v} = 
\left[ \begin{array}{cc} 1&1\\1&2\end{array}\right]^{-1} 
\left[\begin{array}{c} (s+r)+3(r+r_a) \\ (s+2r)+3(r+2r_a) \end{array}\right] = 
\left[\begin{array}{c} s+3r \\ r+3r_a \end{array}\right].
\]
A similar procedure is executed at participants $4$, $5$ and $6$ as well.
\end{example}

\subsection{Correctness of the Algorithm}\label{sec:algorithm_proofs}
The following theorems show that each participant indeed receives its intended share~\eqref{eq:share}, and the algorithm satisfies the properties of $k$-secret-recovery, and $(k-1)$-collusion-resistance, and that the $(k-1)$-collusion-resistance property is also robust to network structure.

\begin{theorem}[Successful share dissemination]\label{thm:disseminate}
Under the algorithm presented, every participant $\ell \in [n]$ can recover $\boldsymbol{\psi}_\ell^T M$, and hence obtain its intended share \[ \boldsymbol{t}_{\ell}^T = \boldsymbol{\psi}_\ell^T \left[\begin{array}{cc}\Msa&\Msc^T\\\Mrb&\Mre^T \\\Msc&0~\end{array}\right]~. \]
\end{theorem}
\begin{IEEEproof}
Recall that the graph satisfies the $d$-dealer propagation condition. Let us assume without loss of generality that the ordering of nodes satisfying this condition is $1,\ldots,n$. It follows that the first $d$ nodes in this ordering must be connected directly to the dealer. The proof proceeds via induction. The induction hypothesis is as follows: every participant $\ell$ can recover the data $\boldsymbol{\psi}_{\ell}^T M$, and if $\ell$ passes any data to any other node $j \in \mathcal{N}(\ell)$ then this data is precisely the value $\boldsymbol{\psi}_{\ell}^T M \boldsymbol{\psi}_j$. 

Consider the base case of node $1$. Since this node is directly connected to the dealer, it receives the data $\boldsymbol{\psi}_1^T M$ from the dealer. Moreover, following the communication protocol, it passes $\boldsymbol{\psi}_1^T M \boldsymbol{\psi}_j$ to each neighbour $j \in \mathcal{N}(1)$. Let us now assume that the hypothesis holds true for the first $(\ell-1)$ nodes in the ordering. If node $\ell$ is directly connected to the dealer, then the hypothesis is satisfied for this node by an argument identical to the case of node $1$. Suppose $\ell$ is not directly connected to the dealer. It follows that node $\ell$ must be connected to at least $d$ other nodes preceding it in the ordering, and furthermore, must receive data from at least $d$ of these nodes (say, nodes $\{j_1,\ldots,j_d\} \subseteq [\ell-1]$). By our hypothesis, these $d$ nodes  pass the $d$ values $\lbrace\boldsymbol{\psi}_{j_1}^TM\boldsymbol{\psi}_\ell,\ldots,\boldsymbol{\psi}_{j_d}^T M \boldsymbol{\psi}_{\ell} \rbrace$. It follows that the algorithm running at node $\ell$ operates on the input
\beq \left[\begin{array}{c}\sigma_1\\ \vdots \\ \sigma_d \end{array} \right] = \left[ \begin{array}{c} ~~~~~~\boldsymbol{\psi}_{j_1}^T~~~~~~ \\\vdots\\\boldsymbol{\psi}_{j_d}^T\end{array}\right] M \boldsymbol{\psi}_\ell~.
\label{eq:recovery_proof1} \eeq
By construction, the matrix in~\eqref{eq:recovery_proof1} with $\boldsymbol{\psi}_{j_1}^T,\ldots,\boldsymbol{\psi}_{j_d}^T$ as its rows is a $(d \times d)$ Vandermonde matrix, and is hence invertible. Thus, the computation of $\boldsymbol{v}$ as described in Algorithm~\ref{algo:comm} can be performed efficiently using standard Reed-Solomon decoding algorithms~\cite{blahut1983theory,gohberg1994fast}. It further follows that $\boldsymbol{v} = M \boldsymbol{\psi}_\ell$, and since $M$ is a symmetric matrix, we get $\boldsymbol{v}^T = \boldsymbol{\psi}_{\ell}^T M^T = \boldsymbol{\psi}_{\ell}^T M$. Finally, the data passed by node $\ell$ to any other node $i \in \mathcal{N}(\ell)$, according to the protocol, is $v^T \boldsymbol{\psi}_i = \boldsymbol{\psi}_{\ell}^T M \boldsymbol{\psi}_i$. This proves the hypothesis for node $\ell$.

Due to the specific structure~\eqref{eq:M} of $M$, the desired share $\boldsymbol{t}_\ell$ is a subset of the elements of the vector $\boldsymbol{\psi}_{\ell}^T M$. Thus, every participant obtains its intended share.
\end{IEEEproof}

\begin{theorem}[$k$-secret-recovery]\label{thm:recovery}
Any $k$ shares suffice to recover the secret.
\end{theorem}
\begin{IEEEproof}
Let $\mathcal{I} \subseteq [n]$ denote the set of the $k$ participants attempting to recover the secret. Let $\Psi_\mathcal{I}$ be a $(k \times d)$ matrix with its $k$ rows comprising $\lbrace \boldsymbol{\psi}_i^T \rbrace_{i\in\mathcal{I}}$. Further, let $\tilde{\Psi}_{\mathcal{I}}$ denote the $(k \times k)$ submatrix of $\Psi_{\mathcal{I}}$ comprising the first $k$ columns of $\Psi_{\mathcal{I}}$. Observe that the $k$ participants in $\mathcal{I}$ collectively have access to the data \[ \Psi_\mathcal{I}\left[\begin{array}{cc}\Msa&\Msc^T\\\Mrb&\Mre^T \\\Msc&0~\end{array}\right].\] Consider the last $(d-k)$ columns of this data, i.e.,\[\Psi_\mathcal{I} \left[\begin{array}{c} \Msc^T \\ \Mre^T \\  0\end{array}\right] = \tilde{\Psi}_\mathcal{I} \left[\begin{array}{c} \Msc^T \\ \Mre^T \end{array}\right].\] Since $\Psi_\mathcal{I}$ is a $(k \times d)$ Vandermonde matrix, it follows that $\tilde{\Psi}_\mathcal{I}$ is a $(k \times k)$ Vandermonde matrix. Thus, $\tilde{\Psi}_\mathcal{I}$ is invertible. This allows for the decoding of $\Msc$ (via algorithms~\cite{blahut1983theory,gohberg1994fast} identical to those for decoding under Shamir's original secret sharing scheme). It now remains to recover $\Msa$, and to this end consider the first column of the data, i.e., \[ \Psi_\mathcal{I}\left[\begin{array}{c}\Msa\\\Mrb\\\Msc~\end{array}\right].\]  Since the value of $\Msc$ is now known, its effect can be subtracted from this data to obtain \[\Psi_\mathcal{I} \left[\begin{array}{c} \Msa \\ \Mrb \\  0\end{array}\right] = \tilde{\Psi}_\mathcal{I} \left[\begin{array}{c} \Msa \\ \Mrb \end{array}\right].\] Since $\tilde{\Psi}_\mathcal{I}$ is invertible, the value of $\Msa$ can be decoded from this data.
\end{IEEEproof}

\begin{theorem}[$(k-1)$-collusion-resistance and robustness]\label{thm:secret}
Any set of $(k-1)$ or fewer colluding participants can gain no information about the secret. This guarantee is robust to network topology, i.e., holds for arbitrary graphs.
\end{theorem}
The proof of Theorem~\ref{thm:secret} is provided in Appendix~\ref{app:proofofsecrecy}.



\begin{remark}
In certain scenarios, the communication network topology may be known beforehand, and it may be desired to verify whether the topology satisfies the $d$-propagating-dealer condition. This task can be performed efficiently by simply simulating the communication protocol of \sneaka (Algorithm~\ref{algo:comm}) on the given network: the $d$-propagating dealer condition is satisfied if and only if all nodes receive data from at least $d$ other nodes. 
\end{remark}

\section{Complexity Analysis and Bounds}\label{sec:analysis}
In this section, we provide an analysis and comparison of the complexity of \sneaka, the \csotas, and lower bounds for any secret-sharing scheme. 


Recall that $D$ denotes the dealer and $\mathcal{N}(D)$ denotes the set of neighbours of the dealer (or, in case of directed edges, the set of nodes with edges coming in from the dealer). Let $|\mathcal{N}(D)|$ denote the size of this set. 
We assume without loss of generality that the units of data are normalized with one unit defined to be equal to the size of the secret. We will use the notation $\Gamma(.)$ to denote communication cost, and $\rho(.)$ to denote amount of randomness required. The proofs of each of the results stated below are available in~Appendix~\ref{app:analysis}.



\subsection{Communication Cost and Randomness Required}
\begin{theorem}\label{thm:complexity_communication_our}
For an $(n,\,k)$ secret sharing problem on any graph $\mathcal{G}$ with $(n+1)$ nodes satisfying the $d$-propagating-dealer condition for some (known) $d$, \sneaka \\
(a) requires every node to receive $\frac{d}{d-k+1}$ units of data, and hence requires a total communication of \beq \Gamma_{\text{\sneak}}(\mathcal{G}) = n\frac{d}{d-k+1} \label{eq:sneak communication}\eeq units of data, and\\
(b) requires an amount of randomness given by \beq \rho_\text{\sneak}(\mathcal{G})  = \frac{(k-1)(2d-k)}{2(d-k+1)}~. \label{eq:sneak randomness} \eeq 
 \label{thm:complexity_our_communication}
\end{theorem}

\begin{theorem}\label{thm:complexity_communication_BGW1}
For an $(n,\,k)$ secret sharing problem on any graph $\mathcal{G}$ with $(n+1)$ nodes, the \csotas  \\
(a) requires a total communication of \beq \Gamma_\sotaabbr(\mathcal{G}) = |\mathcal{N}(D)| + \sum_{i \notin \mathcal{N}(D)} \min_{w \geq k} \left[\frac{w}{w-k+1} \times \ell_w(D \rightarrow i)\right] \label{eq:communication sota}\eeq  units of data, where $\ell_w(D \rightarrow i)$ is the average of the path lengths of the $w$ shortest node-disjoint paths from the dealer to node $i$ (with $\ell_w(D \rightarrow i) = \infty$ if there do not exist $w$ node-disjoint paths from $D$ to $i$), and \\
(b) requires an amount of randomness lower bounded by 
\bea \rho_\sotaabbr(\mathcal{G}) &\geq& k-1 + \sum_{i \notin \mathcal{N}(D)} \frac{k-1}{w_{\text{max}}(i)-k+1} \label{eq:randomness lower 1}\\
&\geq& (n-|\mathcal{N}(D)|)\frac{(k-1)}{|\mathcal{N}(D)|-(k-1)}~\label{eq:randomness sota lower 2}\eea where $w_{max}(i)$ is the maximum number of node-disjoint paths from the dealer to node $i$.
\end{theorem}

\begin{remark}
The lower bound on the randomness requirement of the \csotas provided in~\eqref{eq:randomness lower 1} is achievable, however, at the cost of an increased communication cost (the communication cost will be higher than that specified in~\eqref{eq:communication sota}, wherein the optimal $w$ chosen for every term inside the summation would be replaced by $w_{\text{max}}(i)$).
\end{remark}

From the two theorems stated above, we can see that \sneaka provides the greatest gains over the \csotas when the distance in the graph between the dealer and the participants is large on an average.


The following theorem provides information-theoretic lower bounds under any scheme on the amount of communication and the amount of randomness required, which serves as a benchmark to compare \sneaka and the \csotas.
\begin{theorem}\label{thm:complexity_communication_lower1}
For an $(n,k)$ secret sharing problem on any graph $\mathcal{G}$ with $(n+1)$ nodes under any algorithm, \\
(a) any node $\ell \in [n]$ must receive at least 
\bea \Gamma_\text{every}(\ell) &\geq& 
\begin{cases} 
      \frac{\text{deg}(\ell)}{\text{deg}(\ell)-k +1}  & \text{if~}\ell \notin \mathcal{N}(D) \text{ and } \text{deg}(\ell) \geq k \\
      1 & \text{if~}\ell \in \mathcal{N}(D)\\
      \infty  & \text{if~}\ell \notin \mathcal{N}(D) \text{ and } \text{deg}(\ell) < k
         \end{cases}\nonumber\\&&
         \label{eq:complexity_communication_lower1_eq}
         \eea
units of data, where $\text{deg}(\ell)$ denotes the number of incoming edges at node $\ell$. Furthermore, this bound is the best possible, given only the identities of the neighbours of node $\ell$. Hence the total communication cost under any algorithm is lower bounded by \bea \Gamma_\text{any}(\mathcal{G}) &\geq& |\mathcal{N}(D)| + \sum_{i \notin \mathcal{N}(D)} \frac{\text{deg}(i)}{\text{deg}(i)-k+1}, \label{eq:complexity_communication_lower20_eq} \\
         &\geq& n \label{eq:communication lower3_eq}\eea and \\
(b) the amount of randomness required under any algorithm is lower bounded~\cite{blundo1996randomness} by \beq \rho_\text{any}(\mathcal{G})  \geq k-1~. \eeq
\end{theorem}

\begin{remark}
The lower bound~\eqref{eq:communication lower3_eq} can be deduced alternatively from the fact that the share of each participant must be atleast the size of the secret~[Theorem~$1$,~\cite{karnin1983secret}].
\end{remark}

\subsection{Implications for the Case of Bounded Degree Graphs}
The \sneaka algorithm has particularly striking implications on secret sharing on graphs whose maximum degree is bounded (independent of $n$), for example, in the graph depicted in Fig.~\ref{fig:newtorks_propagating_dealer_layered} with the nodes partitioned into `layers' of three nodes each. 

As discussed earlier, for any secret sharing algorithm to succeed, the graph must satisfy the $d$-connected-dealer condition for some $d \geq k$. Now, if a graph satisfies the $d$-connected-dealer (or the stronger $d$-propagating-dealer) condition, the value of $d$ must be upper bounded by the maximum degree of the graph. It follows that the parameters $k$ and $d$ are upper bounded by the maximum degree of the graph, and are therefore independent of $n$. Theorem~\ref{thm:complexity_communication_main} and Theorem~\ref{thm:randomness bounded degree} present the main results for this setting.

\begin{lemma}\label{thm:complexity_comm_lower4}
For any given $(n,k)$, and for any given $d~(k \leq d <n)$, consider any undirected graph with $(n+1)$ nodes such that (a) every non-neighbour of $D$ has a degree of $d$, and (b) the graph satisfies the $d$-propagating-dealer condition. Under \sneaka, the amount of data received by any node $\ell \notin \mathcal{N}(D)$ meets the lower bound~\eqref{eq:complexity_communication_lower1_eq}. Furthermore, under \sneaka, the amount of data received by any node $\ell \in \mathcal{N}(D)$ is independent of $n$.
\end{lemma}

\begin{lemma}\label{thm:complexity_communication_lower3}
For any given $(n,\, k)$, and for any given $d~(k \leq d <n)$, there exists a class of graphs such that the communication cost on graphs belonging to this class is lower bounded by \beq \Gamma_\text{any}(\mathcal{G}) \geq n\frac{d}{d-k+1} - (k-1)\frac{d}{d-k+1}~. \eeq 
\end{lemma}
Thus, for the class of graphs considered in Lemma~\ref{thm:complexity_communication_lower3}, the communication cost of \sneaka \eqref{eq:sneak communication} is within a constant (additive) factor of the lower bound. 

The following two lemmas quantify the performance of the \csotas. 
Lemma~\ref{thm:complexity_communication_BGW3} is more general than that of bounded degree graphs considered in this section: the lemma also applies to graphs whose maximum degree may grow with $n$.

\begin{lemma}\label{thm:complexity_communication_BGW3}
On graphs with the maximum outgoing degree $O((\text{log\,}n)^{\frac{1}{2}-\epsilon}))$ for some $\epsilon>0$, the \csotas requires a super-linear communication cost. Furthermore, for graphs with degree bounded by a constant independent of $n$, the \csotas requires an $\Omega(n \log n)$ communication cost.
\end{lemma}

\begin{lemma}\label{thm:complexity_communication_BGW2}
For any given $(n,\,k)$, and for any given $d~(k \leq d <n)$, there exists a class of graphs with $(n+1)$ nodes such that each graph in this class satisfies the $d$-propagating dealer condition, and $(n,k)$ secret sharing on any graph $\mathcal{G}$ in this class using the \csotas requires a communication cost lower bounded by \beq \Gamma_\sotaabbr(\mathcal{G}) \geq \frac{n(n+1)}{4d}~. \label{eq:bgw_comm_complexity}\eeq
\end{lemma}

The following theorem gives a comparison between the \csotas, \sneaka and the lower bounds. 
\begin{theorem}\label{thm:complexity_communication_main}
Consider graphs that satisfy the $k$-propagating-dealer condition and have their maximum degree upper bounded by a constant independent of $n$. The  the communication cost of \sneaka is within a constant (multiplicative) factor of the information-theoretic lower bound, and is $\Theta(n)$ in the worst-case. On the other hand, the communication cost of  \csotas is  $\Omega(n\log n)$, and is $\Omega(n^2)$ in the worst case. 
\end{theorem}

The following result quantifies the amount of randomness required for secret sharing under the \csotas and \sneaka.

\begin{theorem}~\label{thm:randomness bounded degree}
Consider graphs that satisfy the $k$-propagating-dealer condition and have their maximum degree upper bounded by a constant independent of $n$. The amount of randomness required under the \csotas is $\Theta(n)$, and the amount of randomness required under \sneaka is $\Theta(1)$.
\end{theorem}


\section{Conclusions and Open Problems}\label{sec:conclusion}
Many cryptographic protocols in the literature require execution of one or more instances of secret sharing among all the participants. Most of these protocols assume that the dealer has direct communication links to all the participants. This paper presents \sneak, a distributed and efficient algorithm for secret sharing in a setting where the dealer and the participants form a general communication network. While \sneak requires the network to satisfy the stronger $k$-propagating-dealer condition as opposed to the $k$-connected-dealer condition required by the existing methods, it provides significant reduction in the communication cost and the amount of randomness required, in addition to being distributed. 
The paper also presents information-theoretic lower bounds on the communication cost for secret sharing in general networks, which may be of independent interest.

The upper and lower bounds on the communication cost for secret sharing presented in this paper are shown to be tight for certain classes of networks. However, obtaining (tight) bounds on the communication cost for general networks still remains open. \sneaka requires the network to satisfy the $k$-propagating dealer condition, and only heuristics are known to address networks that satisfy the $k$-connected-dealer but not the $k$-propagating-dealer condition. Designing more efficient algorithms and quantifying the communication requirements for such settings remain open.

Finally, the results of this paper turn out to be an instance of a network coding problem that interestingly admits an explicit solution that is distributed, communication efficient, and provides probability $1$ guarantees. Moreover, the solution handles the case of nodal eavesdropping, about which very little appears to be known in the literature. As a part of future work, we wish to investigate if any of the ideas from this specific case of secure network coding carry over to more general network coding problems.

\section*{Acknowledgements}
Nihar B. Shah was supported by a Berkeley Fellowship and K. V. Rashmi was supported by a Facebook Fellowship. This work was also supported in part by AFOSR grant FA9550-10-1-0567 and in part by NSF grant CCF-0964018. The authors would like to thank Prakash Ishwar, Piyush Srivastava, Anindya De, and Matthieu Finiasz for helpful discussions.

\bibliographystyle{IEEEtran}
\bibliography{../../bibtex/distributedStorage}


\appendices

\section{Proof of \lowercase{$(k-1)$}-collusion-resistance}\label{app:proofofsecrecy}

\begin{IEEEproof}[Proof of Theorem~\ref{thm:secret}]
This proof follows on the lines of the proof of~\cite[Theorem 1]{ourGlobecom2011}. Consider execution of \sneaka on a network having an arbitrary topology.  Let $\mathcal{J} \subset [n]$ denote the set of $(k-1)$ participants colluding in an attempt to recover information about the secret $\mathbf{s}$. Denote the number of secret values (over $\mathbb{F}_q$) by $S = d-k+1$. Further, let $\mathbf{r}$ denote the collection of (from~\eqref{eq:num_random}) $R = (k-1)d -  {k-1 \choose 2}$ random values introduced initially by the dealer.  

The data obtained by any participant $j \in \mathcal{J}$ is a subset of the values $\psi_j^T M$ and $\lbrace \psi_\ell^T M \psi_j \rbrace_{\ell \in [n]}$. This is true under any arbitrary topology, and irrespective of the connectivity of node $j$. Furthermore, since matrix $M$ is symmetric, $\lbrace \psi_\ell^T M \psi_j \rbrace_{\ell \in [n]} = \lbrace\psi_j^T M \psi_\ell \rbrace_{\ell \in [n]}$. Thus, participant $j$ obtains at most the $d$-length vector $\psi_j^T M$ under the execution of the protocol.

It suffices to consider the worst case wherein every node $j \in \mathcal{J}$ obtains $\psi_j^T M$ completely. Let $\Psi_\mathcal{J}$ be the $((k-1) \times d)$ submatrix of $\Psi$ comprising the $(k-1)$ vectors $\lbrace \boldsymbol{\psi}_j^T \rbrace_{j \in \mathcal{J}}$ as its rows. Under this notation, the $(k-1)$ colluding participants together have access to at most the $(k-1)d$ values  \[ C_\mathcal{J} = \Psi_\mathcal{J} M = \Psi_\mathcal{J} \Mstruct~.\] Let $\mathbf{e}$ denote the set of these $(k-1)d$ values.  

Throughout the proof, we will use the function $H(.)$ to denote the Shannon entropy. All logarithms in the computation of the entropy functions are assumed to be taken to the base $q$.

As an intermediate step in the proof, we will show that given all the secret values $\mathbf{s}$ as side-information, the $(k-1)$ colluding participants can recover all the $R$ random values, i.e., we will show that $H (\mathbf{r} | \mathbf{e}, \mathbf{s})= 0$. To this end, observe that since the code is linear, if the secret values $\Msa$ and $\Msc$ are known to the colluding participants, they can subtract the components of $\Msa$ and $\Msc$ from $C_\mathcal{J}$, to obtain
\[ C^{'}_\mathcal{J} =  \Psi_\mathcal{J} \left[\begin{array}{ccc} 0 & \Mrb^T & 0^T \\ \Mrb & \Mrd & \Mre^T \\0 & \Mre & 0\end{array}\right]~.\]
Since $\Psi_\mathcal{J}$ is a $((k-1)\times d)$ Vandermonde matrix with all non-zero entries, when restricted to columns $2$ to $(k-1)$, it forms a $((k-1)\times(k-1))$ invertible matrix. This allows recovery of the random values in $\Mrb$ and $\Mre$. Subtracting the components of these decoded values gives 
\[ C^{''}_\mathcal{J} = \Psi_\mathcal{J} \left[\begin{array}{ccc} 0 & 0^T & 0^T \\ 0 & \Mrd & 0^T \\0 & 0 & 0\end{array}\right]~,\]
and in a manner identical to that of decoding $\Mrb$ and $\Mre$, the colluding participants can decode the remaining random values $\Mrd$. Thus, given the secret values, the $(k-1)$ colluding participants can decode all the random values, which implies \beq H (\mathbf{r} | \mathbf{e}, \mathbf{s})= 0~.\label{eq:step1}\eeq

As another intermediate step in the proof, we will now show that all but $R$ of the values obtained by the $(k-1)$ participants are functions of the other values that they possess, i.e., $H(\mathbf{e}) \leq R$. From the value of $R$ in~\eqref{eq:num_random}, it suffices to show that out of the $(k-1) d$ values that the colluding participants have access to, ${k-1 \choose 2}$ of them are functions (in particular, linear combinations) of the rest. Consider the $((k-1) \times (k-1))$ matrix
\beq C_\mathcal{J}  \Psi^T_\mathcal{J} = \Psi_\mathcal{J} M\Psi^T_\mathcal{J} ~. \label{eq:mbr_dependent}\eeq
Since $M$ is symmetric, this $((k-1) \times (k-1))$ matrix in~\eqref{eq:mbr_dependent} is also symmetric. Thus $k-1 \choose 2$ dependencies among the elements of $C_\mathcal{J}$ are described by the $k-1 \choose 2$ upper-triangular elements of the expression
\beq  C_\mathcal{J} \Psi^T_\mathcal{J} -\Psi_\mathcal{J} C^T_\mathcal{J} = 0~. \eeq
Since the   rows of $\Psi_\mathcal{J}$ are linearly independent, these ${k-1 \choose 2}$ redundant equations are independent. Thus the colluding participants have access to at most $(k-1) d - {k-1 \choose 2}$ independent values, which equals the value of $R$, and hence\beq H(\mathbf{e}) \leq R. \label{eq:step2}\eeq

We finally show that the two conditions~\eqref{eq:step1} and~\eqref{eq:step2} above must necessarily imply that the mutual information between the secret values and the values obtained by the colluding participants $I(\mathbf{s};\mathbf{e})=0$ is zero.
\begin{eqnarray}
 I(\mathbf{s};\mathbf{e}) &=& H(\mathbf{e}) - H(\mathbf{e} | \mathbf{s}) \label{eq:step3_start} \\
&\leq& R - H(\mathbf{e} |  \mathbf{s})  \label{eq:secrecy_method_1a} \\
&=& R - H(\mathbf{e} |  \mathbf{s}) + H(\mathbf{e} | \mathbf{s},\mathbf{r}) \label{eq:secrecy_method_1b} \\
&=& R - I(\mathbf{e};\mathbf{r} | \mathbf{s}) \\
&=& R - \left( H(\mathbf{r} |  \mathbf{s}) - H(\mathbf{r} |  \mathbf{e},\mathbf{s}) \right) \\
&=& R -  H(\mathbf{r} |  \mathbf{s})  \label{eq:secrecy_method_2} \\
&=& R - R  \label{eq:secrecy_method_3} \\
&=& 0~,\label{eq:step3_stop}\end{eqnarray}
where~\eqref{eq:secrecy_method_1a} follows from~\eqref{eq:step2};~\eqref{eq:secrecy_method_1b} follows since every value in the system is a function of $\mathbf{s}$ and $\mathbf{r}$, giving $H(\mathbf{e} | \mathbf{s},\mathbf{r})=0$;~\eqref{eq:secrecy_method_2} follows from~\eqref{eq:step1}; and~\eqref{eq:secrecy_method_3} follows since the random values are independent of the secret values. Thus, $\mathbf{s}$ and $\mathbf{e}$ are independent random variables.

The $(k-1)$-collusion-resistance property is thus satisfied under any arbitrary topology of the network (that may even vary with time), making \sneaka robust to the network topology.
\end{IEEEproof}

\section{Proofs of Complexity Analysis}\label{app:analysis}
\begin{IEEEproof}[Proof of Theorem~\ref{thm:complexity_communication_our}]
Under \sneaka, each participant who is directly connected to the dealer receives $d$ values (over $\mathbb{F}_q$) and each of the other participant receives exactly one value each from $d$ of its neighbours. Thus, the total amount of data received by each participant is $d$ values. Normalizing this by the size of the secret, which under \sneaka is $(d-k+1)$, gives the desired result~\eqref{eq:sneak communication}.

Under \sneaka, the size of the secret is $(d-k+1)$ values over $\mathbb{F}_q$. From the description of the algorithm~\eqref{eq:num_random} we see that the total randomness required is $\left((k-1)d - {k-1 \choose 2}\right)$ values over $\mathbb{F}_q$. Normalizing this value by the size of the secret gives the desired result~\eqref{eq:sneak randomness}.
\end{IEEEproof}

\def\pathsnum{w}
\begin{IEEEproof}[Proof of Theorem~\ref{thm:complexity_communication_BGW1}]
Under the \csotas, the dealer first computes the (unit-sized) shares $\{t_i\}_{i=1}^{n}$ of each of the nodes. To every node in $\mathcal{N}(D)$, the dealer directly passes its respective share, thus resulting in a total communication of $|\mathcal{N}(D)|$ units of data. Now consider any other node $i \notin \mathcal{N}(D)$. Suppose there exist $\pathsnum~(\geq k)$ node-disjoint paths from the dealer to node~$i$. The dealer treats $t_i$ as a secret and uses two-level secret sharing to construct $\pathsnum$ shares such that no $(k-1)$ of these shares reveal any information about $t_i$, and all $\pathsnum$ shares suffice to recover $t_i$ completely. Under this encoding, the size of each of the $\pathsnum$ shares is $\frac{1}{\pathsnum-k+1}$. The dealer then passes these $\pathsnum$ shares along the $\pathsnum$ node-disjoint paths to node $i$, and node $i$ thus securely obtains $t_i$. Thus, if we let $\ell_\pathsnum(D \rightarrow i)$ denote the average of the lengths of these $\pathsnum$ paths, the communication required is $\frac{\pathsnum}{\pathsnum-k+1} \ell_\pathsnum(D\rightarrow i)$ units of data. The dealer is free to choose the value of $\pathsnum$ across all feasible options, that minimizes this expression, thus leading to the result~\eqref{eq:communication sota}.

Under the \csotas, the dealer first constructs the $n$ shares using $(k-1)$ random values. In addition, in order to disseminate the shares to any node $i \notin \mathcal{N}(D)$, the dealer can perform a two-level secret sharing and pass the shares across $\pathsnum$ node-disjoint paths as described above. This requires $\frac{k-1}{\pathsnum-k+1}$ units of randomness, which is minimized when $\pathsnum = \pathsnum_\text{max}$. This leads to the result~\eqref{eq:randomness lower 1}.

For any graph $\mathcal{G}$ the maximum number of node-disjoint paths from the dealer to any node $i$ is upper bounded by, $\pathsnum_{max}(i) \leq |\mathcal{N}(D)|$. Using this bound in~\eqref{eq:randomness lower 1} leads to the result~\eqref{eq:randomness sota lower 2}.
\end{IEEEproof}

\begin{IEEEproof}[Proof of Theorem~\ref{thm:complexity_communication_lower1}]
Let $S$ denote the random variable representing the secret, and assume without loss of generality that its entropy is normalized to unity, i.e., \beq H(S)=1. \eeq 
\def\degell{z}
Case I ($\ell \notin \mathcal{N}(D)$ and $deg(\ell) \geq k$):  Let $\degell=deg(\ell)$ denote the number of incoming edges at node $\ell$ (in case of an  undirected graph, $z$ denotes the degree of node $\ell$). Let nodes $\{i_1,\ldots,i_\degell\}$ be these $z$ neighbours of node $\ell$. 
 For $j \in \{1,\ldots,\degell\}$, let $W_j$ denote the random variable representing all the data available at node $i_j$, and let $X_j$ denote the random variable representing the data passed by node $i_j$ to node $\ell$. We will now show that
 \beq H(X_1) + \cdots + H(X_{\degell}) \geq \frac{{\degell}}{{\degell}-k+1}, \eeq
which suffices to prove the result.

 Since $W_j$ is the entire data available at node $i_j$, and $X_j$ is the data passed by node $i_j$, it follows that given $W_j$, the random variable $X_j$ is conditionally independent of all other random variables in the set $\{S,X_1,W_1,\ldots,X_{j-1},W_{j-1},X_{j+1},W_{j+1},\ldots,X_{\degell},W_{\degell}\}$. 

Now, for any permutation $\{j_1,\ldots,j_{\degell}\}$ of $\{1,\ldots,{\degell}\}$, it must be that
\begin{align}
H&(X_{j_k}) + \cdots + H(X_{j_{\degell}}) \nonumber\\
&\geq  H(X_{j_k},\ldots, X_{j_{\degell}}) \nonumber\\
&\geq  H(X_{j_k},\ldots, X_{j_{\degell}}|W_{j_1}, \ldots, W_{j_{k-1}}) \nonumber\\
&\geq  I(S;X_{j_k},\ldots, X_{j_{\degell}}|W_{j_1}, \ldots, W_{j_{k-1}}) \nonumber\\
&= H(S|W_{j_1}, ..., W_{j_{k-1}}) \!-\! H(S|W_{j_1}, ..., W_{j_{k-1}}, X_{j_k},..., X_{j_{\degell}})  \nonumber\\
&= 1 - H(S|W_{j_1}, \ldots, W_{j_{k-1}},X_{j_k},\ldots, X_{j_{\degell}}) \label{eq:complexity2}\\
&= 1 - H(S|W_{j_1}, \ldots, W_{j_{k-1}},X_{j_1},\ldots, X_{j_{\degell}}) \label{eq:complexity3}\\
&= 1  \label{eq:complexity4}.
\end{align}
Here,~\eqref{eq:complexity2} is a consequence of the fact that any set of $(k-1)$ nodes cannot contain any information about the secret $S$, and that $H(S)=1$;~\eqref{eq:complexity3} is due to the conditional independence conditions described above;~\eqref{eq:complexity4} arises due to the requirement of being able to recover the secret $S$ from any $k$ nodes.

Now, let us consider the ${\degell}$ cyclic permutations of $\{1,\ldots,{\degell}\}$: $\{1,\ldots,{\degell}\}$, $\{2,\ldots,{\degell},1\}$, $\ldots$, $\{{\degell},1,\ldots,{\degell}-1\}$. Setting $\{j_1,\ldots,j_{\degell}\}$ based on each of these permutations one by one, and adding up the resulting ${\degell}$ inequalities~\eqref{eq:complexity4} we get
\beq ({\degell}-k+1) \left(H(X_1) + \cdots + H(X_{\degell})\right) \geq {\degell}, \eeq
and hence
\beq H(X_1) + \cdots + H(X_{\degell}) \geq \frac{{\degell}}{{\degell}-k+1} \eeq
is the minimum amount of data required to be received at any node that is not a neighbour of the dealer.

Case II ($\ell \in \mathcal{N}(D)$): Let $W_\ell$ denote the entire data received by node $\ell$. Consider a set of some other $(k-1)$ nodes, $\{i_1,\ldots,i_{k-1}\}$. Let $W_{i_1},\ldots,W_{i_{k-1}}$ be the data available at these $(k-1)$ nodes respectively. Then,
\bea H(W_\ell) \!\!&\!\!\geq\!\!&\!\! H(W_\ell | W_{i_1},\ldots,W_{i_{k-1}}) \nonumber \\
\!\!&\!\!\geq\!\!&\!\! I(S;W_\ell | W_{i_1},\ldots,W_{i_{k-1}}) \nonumber \\
\!\!&\!\!=\!\!&\!\! H(S| W_{i_1},\ldots,W_{i_{k-1}}) - H(S|W_\ell,W_{i_1},\ldots,W_{i_{k-1}}) \nonumber \\
\!\!&\!\!=\!\!&\!\!1 
\eea
where the final equation arises from the $k$-secret-recovery and $(k-1)$-collusion-resistance properties, and that we have normalized $H(S)=1$. Thus node $\ell$ receives at least one unit of data.

Case III ($\ell \notin \mathcal{N}(D)$ and $deg(\ell) < k$): It is clear that in this case, $k$-connected-dealer condition is not satisfied, and hence $(n,\,k)$ secret sharing is not possible. This completes the proof of~\eqref{eq:complexity_communication_lower1_eq}. \newline 

We now show that for any $(n,k)$, given only the identities of the neighbours of node $\ell$, this lower bound is the best possible. This is trivially true when $\ell \notin \mathcal{N}(D)$ and $\text{deg}(\ell)<k$, since in this case, the necessary condition of $k$-connected-dealer is not satisfied, thus making $(n,k)$ secret sharing on this graph infeasible. We thus focus on the two remaining cases. Let $\mathcal{A} \subseteq [n]\cup\{D\}\backslash\{\ell\}$ denote the set of nodes that have outgoing edges to node $\ell$, and $\mathcal{B} \subseteq [n]\cup\{D\}\backslash\{\ell\}$ denote the set of nodes which have incoming edges from node $\ell$ (if a node shares an undirected edge with $\ell$, it belongs to both the sets). Given the sets $\mathcal{A}$, $\mathcal{B}$, we will construct a graph on $(n+1)$ nodes consistent with the information about the neighbours of $\ell$     (i.e., the sets $\mathcal{A}$ and $\mathcal{B}$), and show that under \sneaka on this graph, the amount of data required to be received at node $\ell$ matches the bound~\eqref{eq:complexity_communication_lower1_eq} with equality. Consider a graph on $(n+1)$ nodes with the following edge set: an edge from node $\ell$ to every node in $\mathcal{B}$, an edge from every node in $\mathcal{A}$ to node $\ell$, and an edge from $D$ to each node in $[n]\backslash\{\ell\}$. One can verify that this graph is consistent with the information provided about the neighbourhood of node $\ell$. One can also verify that this graph satisfies the $d$-propagating-dealer condition with $d=\text{deg}(\ell)$. In this graph, every node in the set $[n]\backslash\{\ell\}$ receives its share directly from the dealer. If $\ell \in \mathcal{N}(D)$ (i.e., if $D \in \mathcal{A}$), then node $\ell$ also receives its share directly from the dealer. Since the size of the share~\eqref{eq:share} is equal to the size of the secret, the bound~\eqref{eq:complexity_communication_lower1_eq} on the amount of data received is met. When $\ell \notin \mathcal{N}(D)$, Theorem~\ref{thm:complexity_communication_our} shows that under \sneaka with $d=\text{deg}(\ell)$, node $\ell$ obtains its share after receiving exactly $\frac{\text{deg}(\ell)}{\text{deg}(\ell)-k+1}$ units of data. Thus, the lower bound~\eqref{eq:complexity_communication_lower1_eq} is achieved in this case as well.

The result \eqref{eq:complexity_communication_lower20_eq} is an immediate consequence of of the result~\eqref{eq:complexity_communication_lower1_eq} proved above.
\end{IEEEproof}


\begin{IEEEproof}[Proof of Lemma~\ref{thm:complexity_comm_lower4}]
Under \sneaka, each node receives $\left(\frac{d}{d-k+1}\right)$ units of data (as shown in Theorem~\ref{thm:complexity_communication_our}). Since each node $\ell \notin \mathcal{N}(D)$ has $\deg(\ell)=d$, an application of Theorem~\ref{thm:complexity_communication_lower1} leads to the desired result.
\end{IEEEproof}

\begin{IEEEproof}[Proof of Lemma~\ref{thm:complexity_communication_lower3}]
For the given values of $n$, $k$ and $d~(k \leq d <n)$, consider any directed graph with $(n+1)$ nodes such that (a) the dealer has $d$ outgoing edges, (b) every non-neighbour of $D$ has $d$ incoming edges, and (c) the graph satisfies $d$-propagating-dealer condition. Any graph with the above three properties has $|\mathcal{N}(D)|=d$ and $deg(i)=d~~\forall~~i\notin \mathcal{N}(D)$. An application of~\eqref{eq:complexity_communication_lower20_eq} and a simple rearrangement of the terms leads to the desired result.

In order to show that the class of graphs considered above is non-empty, we now present a means to construct graphs with the requisite properties. Consider first, a graph on $(n+1)$ nodes (the dealer and $n$ participants) with no edges. Consider an arbitrary ordering of the $n$ participants as $1,2,\ldots,n$. Add edges in the following manner:
\begin{itemize}
\item For every $i \in \{1,\ldots,d\}$
\begin{itemize}
\item add a directed edge from the dealer to node $i$
\end{itemize}
\item For every $i \in \{d+1,\ldots,n\}$
\begin{itemize}
\item Pick any arbitrary subset $S_i$ of $d$ nodes from the set $\{1,\ldots,i-1\}$
\item For every $j \in S_i$, add a directed edge from node $j$ to node $i$
\end{itemize}
\end{itemize}
This completes the construction of the graph. One can verify that this graph satisfies all the three requisite properties (with the $d$-propagating dealer condition satisfied under the ordering $1,2,\ldots,n$ of the participants). 

While the above description provides a general consturction, a concrete example of a graph satisfying the three properties listed above, is the class of layered graphs as in Fig.~\ref{fig:newtorks_propagating_dealer_layered} with a modification: assume all edges to be directed from the left to the right, and the existence of an additional `dealer' node that has edges to every node in the leftmost layer.
\end{IEEEproof}

\begin{IEEEproof}[Proof of Lemma~\ref{thm:complexity_communication_BGW3}]
For any graph $\mathcal{G}_n$ with $(n+1)$ nodes, let $b_n$ be the maximum of the outgoing degrees of all the nodes in the graph. Since $k>1$, we have $b_n>1$. In~\eqref{eq:communication sota} from Theorem~\ref{thm:complexity_communication_BGW1}(a), observe that
\beq \min_{w>k} \ell_w(D \rightarrow i) \geq \ell_{\text{min}}(D \rightarrow i)\eeq 
where $\ell_{\text{min}}(D \rightarrow i)$ is the length of the shortest path from the dealer to node $i$, and that $\frac{w}{w-k+1} \geq 1, \quad \forall w >k$.
As a consequence of this, we have \beq \Gamma_\sotaabbr (\mathcal{G}_n) \geq |\mathcal{N}(D)| + \sum_{i \notin \mathcal{N}(D)} \ell_{\text{min}}(D \rightarrow i)~. \eeq Now, since the degree of each node is upper bounded by $b_n$, there can be at most $b_n$ nodes that are connected directly to the dealer, at most $b_n^2$ nodes with  $\ell_{\text{min}}(D \rightarrow i)=2$, at most $b_n^3$ nodes with $\ell_{\text{min}}(D \rightarrow i)=3$, and so on. Let 
\bea m &=&\text{arg max } \tilde{m}\label{eq:complexity_communication_BGW3_eq1} \\ &&\text{subject to } \left( \sum_{j=1}^{\tilde{m}} b_n^j \leq n \right)~,\nonumber\\&&~~~~~~~\text{and }\tilde{m} \in \{0,1,2,\ldots\}.\nonumber\eea
It follows from the discussion above that
\bea \Gamma_\sotaabbr (\mathcal{G}_n) &\geq& \sum_{j=1}^{m} j b_n^j\\ &\geq& m b_n^m~. \label{eq:complexity_communication_BGW3_eq2}\eea
One can infer from~\eqref{eq:complexity_communication_BGW3_eq1} that $m$ is the largest integer satisfying
\[ \frac{b_n^{m+1}-b_n}{b_n-1} \leq n~,\]
and hence
\begin{align} m &= \left\lfloor \frac{\log(n(b_n-1))+b_n}{\log b_n} \right\rfloor -1 \nonumber \\
& \geq \frac{\text{log\,}n}{\text{log\,}b_n}-2~.
\end{align}
Substituting this in~\eqref{eq:complexity_communication_BGW3_eq2}, we obtain
\bea \Gamma_\sotaabbr (\mathcal{G}_n) &\geq&  \left(\frac{\text{log\,}n}{\text{log\,}b_n}-2\right)\frac{n}{b_n^2} \\
\Rightarrow \frac{\Gamma_\sotaabbr (\mathcal{G}_n)}{n} &\geq& \left(\frac{\text{log\,}n}{\text{log\,}b_n}-2\right)\frac{1}{b_n^2}~. \eea
Setting $b_n = O((\text{log\,}n)^{\frac{1}{2}-\epsilon}))$ or $b_n = O(1)$ gives the respective desired results. 
\end{IEEEproof}

\begin{IEEEproof}[Proof of Lemma~\ref{thm:complexity_communication_BGW2}]
For any given $(n,k)$ and $d~(k \leq d <n)$, consider a parameter $a$ where $0 \leq a \leq n$. Construct a graph $\mathcal{G}$ with $(n+1)$ nodes in the following manner. Consider first a graph on $(n+1)$ nodes (the dealer and $n$ participants) with no edges. Consider an arbitrary ordering of the $n$ participants as $1,2,\ldots,n$. Denote the delaer as node $0$. Add edges in the following manner:
\begin{itemize}
\item For every $i \in \{1,\ldots,d\}$
\begin{itemize}
\item add a directed edge from the dealer to node $i$
\end{itemize}
\item For every $i \in \{d+1,\ldots,n\}$
\begin{itemize}
\item Pick any arbitrary subset $S_i$ of $d$ nodes from the set $\{\max(i-(d+a),0),\ldots,i-1\}$
\item For every $j \in S_i$, add a directed edge from node $j$ to node $i$
\end{itemize}
\end{itemize}
From the construction described above, it is clear that each graph in this class satisfies the $d$-propagating-dealer condition (with the requisite ordering of the nodes being $1,\ldots,n$). 

Clearly under this construction, any path from the dealer to any node $i$ requires at least $\left\lceil \frac{i}{d+a} \right\rceil$ steps. Now, the expression in the statement of Theorem~\ref{thm:complexity_communication_BGW1} can be bounded as
\beq \Gamma_\sotaabbr (\mathcal{G}) \geq |\mathcal{N}(D)| + \sum_{i \notin \mathcal{N}(D)} \ell_{\text{min}}(D \rightarrow i) \eeq where $\ell_{\text{min}}(D \rightarrow i)$ is the length of the shortest path from the dealer to node $i$. This can be rewritten in the present context as
\bea \Gamma_\sotaabbr (\mathcal{G}) &\geq& \sum_{i=1}^{n} \left\lceil \frac{i}{d+a} \right\rceil  \\
&\geq& \frac{n(n+1)}{2(d+a)}. \label{eq:complexity_communication_BGW2_1}
\eea
Setting $a=d$ in~\eqref{eq:complexity_communication_BGW2_1} leads to the desired result.

\end{IEEEproof}

\begin{IEEEproof}[Proof of Theorem~\ref{thm:complexity_communication_main}]
A consequence of Proposition~\ref{prop:necessary} is that for a graph having a maximum degree that is upper bounded by a constant independent of $n$, the parameters $k$ and $d$ must also be bounded by that constant, and hence are $O(1)$. The result now follows from Theorem~\ref{thm:complexity_communication_our}(a), Theorem~\ref{thm:complexity_communication_BGW1}(a), Lemma~\ref{thm:complexity_communication_BGW3}, and Lemma~\ref{thm:complexity_communication_BGW2}.
\end{IEEEproof}

\begin{IEEEproof}[Proof of Theorem~\ref{thm:randomness bounded degree}]
From ~\eqref{eq:randomness sota lower 2} we can see that the amount of randomness required under the \csotas increases with $n$ unless the number of nodes connected directly to the dealer also increases linearly with $n$. A consequence of Proposition~\ref{prop:necessary} is that for a graph having a maximum degree that is upper bounded by a constant independent of $n$, the parameters $k$ and $d$ must also be bounded by that constant, and hence are $O(1)$. Furthermore, since the maximum degree is upper bounded by a constant independent of $n$, so is the number of neighbours of the dealer. Then the result follows from Theorem~\ref{thm:complexity_communication_our} and Theorem~\ref{thm:complexity_communication_BGW1}
\end{IEEEproof}

\section{Extensions of the \sneak Algorithm}\label{app:extensions}
\subsection{Two-threshold Secret Sharing Over General Networks}
In~\cite{franklin1992communication}, the authors introduced a modification of Shamir's secret sharing scheme to include two thresholds $k$ and $\ell~(<k)$. The modified scheme satisfies the properties of $k$-secret-recovery and $\ell$-collusion-resistance (Shamir's original scheme is a special case with $\ell = k-1$).  The relaxation of $\ell$ to a value smaller than $(k-1)$ allows for the reduction of the size of each share (when normalized by the message size), thus requiring the dealer to transmit a smaller amount of data, and the participants to store lesser data.

We now generalize the \sneaka algorithm presented in Section~\ref{sec:algorithm}, for distributed secret sharing across a general network, to accommodate two thresholds. The generalization only modifies the structure of matrix $M$ in~\eqref{eq:M} in the original algorithm. Given two thresholds $k$ and $\ell$, the dimensions of the constituent submatrices of $M$ are changed to
\def\Msa{S_A}
\def\Mrb{R_a}
\def\Msc{S_B}
\def\Mrd{R_b}
\def\Mre{R_c}
\bea M &=& \Mstruct \label{eq:M2} \\
&& ~~\underbrace{\!\!\!\!\!}_{k-\ell}\,\underbrace{~~~~~~}_{\ell}\,\underbrace{~~~~~~~~}_{d-k}  \nonumber\\
&& ~~\underbrace{~~~~~~~~~~~~~~~~~~~~~}_{d} \nonumber
\eea
where
 \begin{itemize}
\item $\Msa$ is a \textit{symmetric} $((k-\ell) \times (k-\ell))$ matrix containing $\frac{(k-\ell)(k-\ell+1)}{2}$ secret values,
\item $\Msc$ is $((d-k) \times (k-\ell)   )$ matrix containing $(k-\ell)(d-k)$ secret values,
\item $\Mrb$ is a $(\ell \times (k-\ell))$ matrix containing $\ell(k-\ell)$ random values,
\item $\Mrd$ is a $(\ell\times \ell)$ \textit{symmetric} matrix containing $\frac{\ell(\ell+1)}{2}$ random values,
\item $\Mre$ is a $((d-k) \times \ell)$ matrix containing $\ell(d-k)$ random values.
\end{itemize}
Each random or secret value is drawn from the finite field $\mathbb{F}_q$, $q >n$. Note that $M$ continues to be a $(d \times d)$ symmetric matrix. The remaining  algorithm remains the same as in Section~\ref{sec:algorithm}. The properties of $k$-secret-recovery, $\ell$-collusion-resistance and robustness to network structure can be verified via arguments analogous to those in Section~\ref{sec:algorithm_proofs}.

\subsection{Handling Active Adversaries}\label{app:extensions_active}
Throughout the paper we assumed a honest-but-curious model, where the participants honestly follow the protocol, but may gather any available information. Now, we consider the case when some participants may be active adversaries, i.e., may pass corrupt values to their neighbours (in addition to trying to gather information about the secret). We show how to modify \sneaka (Section~\ref{sec:algorithm}) to handle the case when there are upto $t$ active adversaries in the system, for some given parameter $t$.

The modified algorithm requires the network to satisfy a $(d+2t)$-propagating-dealer condition; let us assume this holds. Under the algorithm, the dealer computes the matrix $M$ and the encoding vectors as described in Section~\ref{sec:algorithm}. As before, to each participant $i$ directly connected to the dealer, it passes the data $\boldsymbol{\psi}_i^T M$. The only modifications are that each participant $\ell$ who is not connected to the dealer obtains data from $(d+2t)$ neighbours and that the method of recovering the data $\boldsymbol{\psi}_\ell^T M$ now involves correcting errors. 
Let us assume that participant $\ell$ receives data form neighbours $\{j_1,\ldots,j_{d+2t}\}$. According to the protocol, this data is the set of $(d+2t)$ values $\{ \boldsymbol{\psi}_{j_1}^T M \boldsymbol{\psi}_\ell,\ldots,\boldsymbol{\psi}_{j_{d+2t}}^T M \boldsymbol{\psi}_\ell\}$. By construction, any $d$ vectors from the set $\{ \boldsymbol{\psi}_{j_1}^T ,\ldots,\boldsymbol{\psi}_{j_{d+2t}}^T\}$ are linearly independent. Thus, the $(d+2t)$ values received by node $\ell$ form a Maximum-Distance-Separable (MDS) encoding of the $d$-length vector $M \boldsymbol{\psi}_\ell$. Furthermore, since at most $t$ of the participants may be actively adversarial, no more than $t$ out of the $(d+2t)$ received values can be in error. Thus, participant $\ell$ can apply standard Reed-Solomon code decoding algorithms~\cite{blahut1983theory} and recover $M \boldsymbol{\psi}_\ell$ correctly. Finally, since $M$ is symmetric by construction~\eqref{eq:M}, participant $\ell$ equivalently obtains its desired data $\boldsymbol{\psi}_\ell^T M$. The participant then passes $\boldsymbol{\psi}_\ell^T M \boldsymbol{\psi}_j$ to each of its neighbours $j$.

\subsection{Heuristics for Handling Networks that Do Not Meet the Propagating-dealer Condition}\label{app:extension}
\newsavebox{\appendixnetwork}
\def\figToyHeight{1.1in}
\def\figToyHeightC{2in}
\def\needWidth{4.85in}
\def\needHeight{1.52in}
\def\nodeDia{11pt}
\def\nodegap{2.8}
\def\nodeAx{.2}
\def\nodeAy{6.7}

\def\nodeBx{\nodeAx}
\def\nodeBy{\nodeAy-\nodegap}
\def\nodeCx{\nodeAx+.8*\nodegap}
\def\nodeCy{\nodeAy}
\def\nodeDx{\nodeBx+.8*\nodegap}
\def\nodeDy{\nodeBy}
\def\nodeEx{\nodeCx+.8*\nodegap}
\def\nodeEy{\nodeAy}
\def\nodeFx{\nodeDx+.8*\nodegap}
\def\nodeFy{\nodeBy}
\def\nodeGx{\nodeEx+.8*\nodegap}
\def\nodeGy{\nodeAy}
\def\nodeHx{\nodeFx+.8*\nodegap}
\def\nodeHy{\nodeBy}
\def\nodeIx{\nodeGx+.8*\nodegap}
\def\nodeIy{\nodeAy}
\def\nodeJx{\nodeHx+.8*\nodegap}
\def\nodeJy{\nodeBy}
\def\nodeKx{\nodeIx+.8*\nodegap}
\def\nodeKy{\nodeAy}
\def\nodeLx{\nodeJx+.8*\nodegap}
\def\nodeLy{\nodeBy}
\def\nodeMx{\nodeKx+.8*\nodegap}
\def\nodeMy{\nodeAy}
\def\nodeNx{\nodeLx+.8*\nodegap}
\def\nodeNy{\nodeBy}
\def\nodeDealerwidth{1.5*\nodeDia}
\def\nodeDealerheight{1.5*\nodeDia}
\def\nodeDealerx{\nodeAx-.6*\nodegap-.5*\nodeDealerwidth*2.5/72} 
\def\nodeDealery{\nodeBy+.65*\nodegap-.5*\nodeDealerheight*2.5/72)}
\def\nodeDealercornerx{\nodeDealerx-.5*\nodeDealerwidth*2.5/72}
\def\nodeDealercornery{\nodeDealery-.5*\nodeDealerheight*2.5/72}
\def\textedgeoffset{7pt}
\def\textedgeoffsetdown{-6pt}
\def\textedgeoffsetleft{9pt}
\def\textedgeoffsetleftdown{-6pt}
\def\textedgeoffsetright{6pt}
\def\textedgeoffsetrightdown{-9pt}

\def\vspacingBetnGraphs{.7cm}

\savebox{\appendixnetwork}{
\begin{pgfpicture}{0cm}{0cm}{\needWidth}{\needHeight}
\pgfline{\pgfxy(\nodeAx,\nodeAy)}{\pgfxy(\nodeCx,\nodeCy)}
\pgfline{\pgfxy(\nodeAx,\nodeAy)}{\pgfxy(\nodeDx,\nodeDy)}
\pgfline{\pgfxy(\nodeBx,\nodeBy)}{\pgfxy(\nodeDx,\nodeDy)}
\pgfline{\pgfxy(\nodeCx,\nodeCy)}{\pgfxy(\nodeEx,\nodeEy)}
\pgfline{\pgfxy(\nodeDx,\nodeDy)}{\pgfxy(\nodeFx,\nodeFy)}
\pgfline{\pgfxy(\nodeBx,\nodeBy)}{\pgfxy(\nodeCx,\nodeCy)}
\pgfline{\pgfxy(\nodeGx,\nodeGy)}{\pgfxy(\nodeJx,\nodeJy)}
\pgfline{\pgfxy(\nodeDx,\nodeDy)}{\pgfxy(\nodeEx,\nodeEy)}
\pgfline{\pgfxy(\nodeCx,\nodeCy)}{\pgfxy(\nodeFx,\nodeFy)}
\pgfline{\pgfxy(\nodeHx,\nodeHy)}{\pgfxy(\nodeIx,\nodeIy)}
\pgfline{\pgfxy(\nodeEx,\nodeEy)}{\pgfxy(\nodeGx,\nodeGy)}
\pgfline{\pgfxy(\nodeEx,\nodeEy)}{\pgfxy(\nodeHx,\nodeHy)}
\pgfline{\pgfxy(\nodeFx,\nodeFy)}{\pgfxy(\nodeGx,\nodeGy)}
\pgfline{\pgfxy(\nodeGx,\nodeGy)}{\pgfxy(\nodeIx,\nodeIy)}
\pgfline{\pgfxy(\nodeHx,\nodeHy)}{\pgfxy(\nodeJx,\nodeJy)}
\pgfline{\pgfxy(\nodeDealerx,\nodeDealery)}{\pgfxy(\nodeAx,\nodeAy)}
\pgfline{\pgfxy(\nodeDealerx,\nodeDealery)}{\pgfxy(\nodeBx,\nodeBy)}

\color{\colornode}
\pgfcircle[fill]{\pgfxy(\nodeAx,\nodeAy)}{\nodeDia}
\pgfcircle[fill]{\pgfxy(\nodeBx,\nodeBy)}{\nodeDia}
\pgfcircle[fill]{\pgfxy(\nodeCx,\nodeCy)}{\nodeDia}
\pgfcircle[fill]{\pgfxy(\nodeDx,\nodeDy)}{\nodeDia}
\pgfcircle[fill]{\pgfxy(\nodeEx,\nodeEy)}{\nodeDia}
\pgfcircle[fill]{\pgfxy(\nodeFx,\nodeFy)}{\nodeDia}
\pgfcircle[fill]{\pgfxy(\nodeGx,\nodeGy)}{\nodeDia}
\pgfcircle[fill]{\pgfxy(\nodeHx,\nodeHy)}{\nodeDia}
\pgfcircle[fill]{\pgfxy(\nodeIx,\nodeIy)}{\nodeDia}
\pgfcircle[fill]{\pgfxy(\nodeJx,\nodeJy)}{\nodeDia}
\color{\colordealer}
\pgfrect[fill]{\pgfxy(\nodeDealercornerx,\nodeDealercornery)}{\pgfpoint{\nodeDealerwidth}{\nodeDealerheight}}
\color{black}
\pgfcircle[stroke]{\pgfxy(\nodeAx,\nodeAy)}{\nodeDia}
\pgfcircle[stroke]{\pgfxy(\nodeBx,\nodeBy)}{\nodeDia}
\pgfcircle[stroke]{\pgfxy(\nodeCx,\nodeCy)}{\nodeDia}
\pgfcircle[stroke]{\pgfxy(\nodeDx,\nodeDy)}{\nodeDia}
\pgfcircle[stroke]{\pgfxy(\nodeEx,\nodeEy)}{\nodeDia}
\pgfcircle[stroke]{\pgfxy(\nodeFx,\nodeFy)}{\nodeDia}
\pgfcircle[stroke]{\pgfxy(\nodeGx,\nodeGy)}{\nodeDia}
\pgfcircle[stroke]{\pgfxy(\nodeHx,\nodeHy)}{\nodeDia}
\pgfcircle[stroke]{\pgfxy(\nodeIx,\nodeIy)}{\nodeDia}
\pgfcircle[stroke]{\pgfxy(\nodeJx,\nodeJy)}{\nodeDia}
\pgfrect[stroke]{\pgfxy(\nodeDealercornerx,\nodeDealercornery)}{\pgfpoint{\nodeDealerwidth}{\nodeDealerheight}}

\pgfputat{\pgfxy(\nodeAx,\nodeAy)}{\pgfbox[center,center]{\Large 1}}
\pgfputat{\pgfxy(\nodeBx,\nodeBy)}{\pgfbox[center,center]{\Large2}}
\pgfputat{\pgfxy(\nodeCx,\nodeCy)}{\pgfbox[center,center]{\Large3}}
\pgfputat{\pgfxy(\nodeDx,\nodeDy)}{\pgfbox[center,center]{\Large4}}
\pgfputat{\pgfxy(\nodeEx,\nodeEy)}{\pgfbox[center,center]{\Large5}}
\pgfputat{\pgfxy(\nodeFx,\nodeFy)}{\pgfbox[center,center]{\Large6}}
\pgfputat{\pgfxy(\nodeGx,\nodeGy)}{\pgfbox[center,center]{\Large7}}
\pgfputat{\pgfxy(\nodeHx,\nodeHy)}{\pgfbox[center,center]{\Large8}}
\pgfputat{\pgfxy(\nodeIx,\nodeIy)}{\pgfbox[center,center]{\Large9}}
\pgfputat{\pgfxy(\nodeJx,\nodeJy)}{\pgfbox[center,center]{\Large10}}
\pgfputat{\pgfxy(\nodeDealerx,\nodeDealery)}{\pgfbox[center,center]{\Large D}}
\end{pgfpicture}
}

\begin{figure*}[b!]
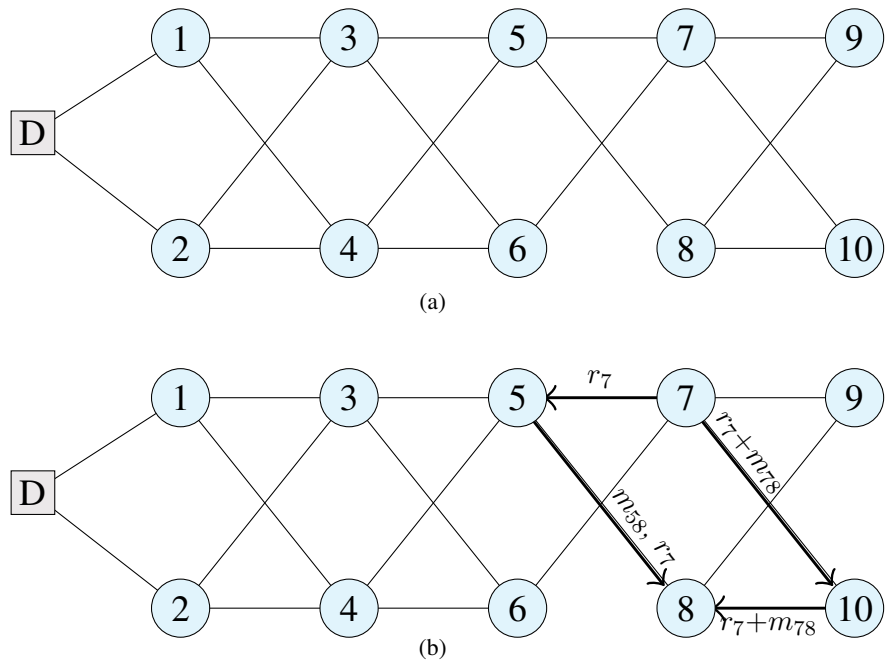

\medmuskip=0\medmuskip
\begin{minipage}[b]{.69\textwidth}
\vspace{3.2cm}
\subfloat[]{
\hspace{5cm}
\begin{minipage}[l]{\textwidth}
\begin{pgfpicture}{0}{0}{\needWidth}{\needHeight}
\usebox{\appendixnetwork}
\end{pgfpicture}
\vspace{-3.5cm}
\end{minipage}
\label{fig:app_assum_1}
}
\end{minipage}
\vspace{3.9cm}
\\
\begin{minipage}[b]{.69\textwidth}
\subfloat[]{
\hspace{5cm}
\begin{minipage}[l]{\textwidth}
\begin{pgfpicture}{0}{0}{\needWidth}{\needHeight}
\def\patha{$r_7+m_{78}$}
\def\pathb{$r_7$}
\def\pathc{$\!\!\!m_{58},\,r_7$}
\def\shiftx{.1pt}
\def\arrowshiftx{.28\nodeDia}
\def\arrowshifty{.33\nodeDia}
\pgfsetlinewidth{1.1pt}
\pgfsetendarrow{\pgfarrowto}
\pgfline{\pgfxy(\nodeGx+\shiftx,\nodeGy)}{\pgfxy(\nodeJx+\shiftx-\arrowshiftx,\nodeJy+\arrowshifty)}
\pgfsetstartarrow{\pgfarrowto}
\pgfsetendarrow{\pgfarrowto}
\pgfline{\pgfxy(\nodeEx+\shiftx,\nodeEy)}{\pgfxy(\nodeHx+\shiftx-\arrowshiftx,\nodeHy+\arrowshifty)}
\pgfline{\pgfxy(\nodeGx+\shiftx,\nodeGy)}{\pgfxy(\nodeEx+\shiftx+.42\nodeDia,\nodeEy)}
\pgfsetstartarrow{\pgfarrowto}
\pgfline{\pgfxy(\nodeHx+\shiftx+.42\nodeDia,\nodeHy)}{\pgfxy(\nodeJx,\nodeJy)}

\pgfputlabelrotated{.33}{\pgfxy(\nodeGx,\nodeGy)}{\pgfxy(\nodeJx,\nodeJy)}{\textedgeoffsetleft}{\pgfbox[center,center]{\patha}}
\pgfputlabelrotated{.55}{\pgfxy(\nodeEx,\nodeEy)}{\pgfxy(\nodeGx,\nodeGy)}{\textedgeoffset}{\pgfbox[center,center]{\pathb}}
\pgfputlabelrotated{.55}{\pgfxy(\nodeHx,\nodeHy)}{\pgfxy(\nodeJx,\nodeJy)}{\textedgeoffsetdown}{\pgfbox[center,center]{\patha}}
\pgfputlabelrotated{.72}{\pgfxy(\nodeEx,\nodeEy)}{\pgfxy(\nodeHx,\nodeHy)}{\textedgeoffsetleft}{\pgfbox[center,center]{\pathc}}
\usebox{\appendixnetwork}
\end{pgfpicture}
\vspace{-3.7cm}
\end{minipage}
\label{fig:app_assum_2}
}
\end{minipage}
\caption{(a) An example network that does not satisfy the $k$-propagating dealer condition for $k=2$. Node $8$ is a bottleneck node. (b) Communicating data to the bottleneck node $8$: node $7$ communicates its message $m_{78}=((s+7r)+8(r+7r_a))$ to node $8$ under \sneaka, through the two highlighted node-disjoint paths ($r_7$ is chosen uniformly at random from the field of operation $\mathbb{F}_{11}$). Node $8$ also obtains $m_{58}=((s+5r)+8(r+5r_a))$ directly from node $5$.}
\end{figure*}
\sneaka requires the graph to satisfy the $d$-propagating-dealer condition for some known $d~(\geq k)$. If a graph does not satisfy the $d$-propagating-dealer condition, there will exist a subset of the nodes that will not be able to recover their respective shares~\footnote{Recall that no information will, however, be leaked.}. For example, consider $(n=10,\,k=2)$ secret-sharing over the graph shown in Fig.~\ref{fig:app_assum_1} using \sneaka with parameter $d=2$. In this example, the graph does not satisfy $2$-propagating-dealer condition since only one neighbour of node $8$ (which is node $5$) can recover the data prior to node $8$. If \sneaka is used to disseminate the shares assuming that the graph satisfies $2$-propagating dealer condition,  nodes $8,\,9,\, \text{and} \,10$ will not be able to recover their shares. 

We now present three heuristic ways of extending \sneaka to handle secret sharing over graphs which do not satisfy the $d$-propagating-dealer condition, using the graph in Fig.~\ref{fig:app_assum_1} as a working example. We note that while these heuristics work successfully, the resulting algorithm is no longer completely distributed and a rigorous analysis of its performance guarantees is open.

\paragraph{Heuristic~$1$}
One straightforward approach is to employ the \csotas of separate secure transmissions from the dealer to all the nodes that do not receive their  respective shares upon an execution of \sneaka. Using this approach for the example in Fig.~\ref{fig:app_assum_1}, for each of the nodes $i \in \{8,\,9,\,10\}$, the dealer communicates the respective share $(s+ir)$ over two node-disjoint paths, using the \csotas solution. This requires a total of $30$ units of communication to disseminate shares to these nodes. In general, this approach is inefficient, since it does not exploit the advantages offered by our distributed algorithm to the fullest extent. This is especially so when only a few nodes act as \textit{bottlenecks} hindering the progress of the algorithm as we will see below. 

\paragraph{Heuristic~$2$}
Observe that in the example network in Fig.~\ref{fig:app_assum_1} there is only one node, node $8$, that is the bottleneck: receipt of its data by node $8$ would allow \sneaka to continue further and disseminate shares to the remaining nodes ($9$ and $10$) efficiently. This leads to a second approach that is more communication efficient, wherein the \csotas can be employed to communicate data to only the bottleneck nodes. The \csotas would pass precisely the data that the bottleneck node(s) would possess under \sneaka, after which \sneaka can be employed to disseminate shares to the remaining nodes. For the example under consideration, node $i \in \{1,\ldots,7\}$ would have obtained data $\{s+ir,r+ir_a\}$ via \sneaka. Next, the dealer would communicate $\{s+8r,r+8r_a\}$ to node $8$ through the two node-disjoint paths, $1 \rightarrow 3 \rightarrow 5 \rightarrow 8$ and $ 2 \rightarrow 4 \rightarrow 6 \rightarrow 7 \rightarrow 9 \rightarrow 8$, and subsequently, nodes $7$ and $8$ can pass the requisite shares to the remaining nodes $9$ and $10$. This approach requires a total of $14$ units of communication to disseminate shares to the nodes $8,\,9\,\text{and } 10$. Note that the graph considered in the example had only one bottleneck node, and hence the \csotas was employed only once. On a general graph, this approach can be iteratively  performed (whenever \sneaka hits a bottleneck node) until all the participants receive their shares. Further note that the dealer can always communicate the data to the bottleneck nodes since the graph has to \textit{necessarily} satisfy the $k$-connected-dealer condition in order to achieve $(n, \, k)$ secret sharing.

\paragraph{Heuristic~$3$}
Instead of communicating the data to the bottleneck nodes from the dealer, one may alternatively use a \textit{local} version of the \csotas: nodes in the vicinity of the bottleneck nodes, who have already received their data, pass the requisite data to the bottleneck nodes via node-disjoint paths. The data that such a node passes to the bottleneck node is precisely what it would have passed had there been a direct edge between them using the \csotas. This node treats this data as a secret, and uses $k$-node-disjoint paths (if available) to communicate this secret to the bottleneck node. To illustrate this approach, consider again the network depicted in Fig.~\ref{fig:app_assum_1}. Node $8$ can directly receive $m_{58}=((s+5r)+8(r+5r_a))$ from node $5$. Now, if node $7$ had a direct communication link to node $8$, under \sneaka, it would have sent  $m_{78}=((s+7r)+8(r+7r_a))$ to node $8$. In the absence of such an edge, node $7$ can securely communicate this data over two node-disjoint paths as depicted in Fig.~\ref{fig:app_assum_2}. Once node $8$ gets this data, \sneaka can proceed and disseminate shares to the remaining nodes $9$ and $10$. This approach requires only $8$ units of communication to disseminate shares to nodes $8,\,9\,\text{and } 10$. This simple tweak can be employed to reduce the communication cost to the bottleneck nodes, whenever sufficient local connectivity is available.

\end{document}